\documentclass[10pt]{iopart}
\usepackage{graphicx, bm}
\usepackage{cite}
\usepackage{braket}
\usepackage{flushend}
\begin{document}

\title{Phase matching in $\beta$-barium borate crystals for spontaneous parametric down-conversion}
\author{Suman Karan$^1$, Shaurya Aarav$^{1}$\footnote{Present address: Department of Electrical Engg., Princeton University, Princeton, NJ 08544, USA.}, Homanga Bharadhwaj$^2$\footnote{Present address: Department of Computer Sc., University of Toronto \& Vector Institute,Canada.}, Lavanya Taneja $^1$\footnote{Present address: Department of Physics,
University of Chicago, Chicago, IL 60637, USA.}, Arinjoy De$^3$\footnote{Present address:Department of Physics, University of Maryland, College Park, MD 20742, USA.}, Girish Kulkarni$^1$\footnote{Present address: Department of Physics, University of Ottawa, Ottawa, ON K1N6N5, Canada.}, Nilakantha Meher$^1$, and Anand K Jha$^{1,*}$}

\address{$^1$ Department of Physics, Indian Institute of Technology Kanpur, Kanpur UP 208016, India.}
\address{$^2$ Department of Computer Science \& Engg., IIT Kanpur, 208016, India.}
\address{$^3$ Department of Physics, IIT Kharagpur, Kharagpur 721302, India.}
\ead{$^*$akjha9@gmail.com}

\begin{abstract}\\

Spontaneous parametric down-conversion (SPDC) is the most widely used process for generating photon pairs entangled in various degrees of freedom such as polarization, time-energy, position-transverse momentum, and angle-orbital angular momentum (OAM). In SPDC, a pump photon interacts with a nonlinear optical crystal and splits into two entangled photons called the signal and the idler photons. The SPDC process has been studied extensively in the last few decades for various pump and crystal configurations, and the entangled photon pairs produced by SPDC have been used in numerous experimental studies on quantum entanglement and entanglement-based real-world quantum-information applications. In this tutorial article, we present a thorough study of phase matching in $\beta$-barium borate (BBO) crystals for spontaneous parametric down-conversion and thereby also investigate the generation of entangled photons in such crystals. First, we present a theoretical derivation of two-photon wavefunction produced by SPDC in the frequency and transverse momentum bases. We then discuss in detail the effects due to various crystal and pump parameters  including the length of the crystal, the angle between the optic axis and the pump propagation direction, the pump incidence angle on the crystal surface, the refraction at the crystal surfaces, and the pump propagation direction inside the crystal. These effects are extremely  relevant in experimental situations. We then present our numerical and experimental results in order to illustrate how various experimental parameters affect the phase matching and thus the generation of entangled photons. Finally, using the two-photon wavefunction in the transverse wave-vector basis, we show how to derive the two-photon wavefunction in the OAM basis and thereby calculate the two-photon angular Schmidt spectrum. We expect this article to be useful for researchers working in various capacities with entangled photons generated by SPDC in BBO crystals.   

\end{abstract}
\noindent{\it Keywords : Parametric down-conversion, Phase matching, Entanglement, Orbital angular momentum }

\ioptwocol
\tableofcontents

\section{Introduction} \label{intro}

\subsection{Background}\label{background}

Spontaneous parametric down-conversion (SPDC) is a second-order nonlinear optical process in which a photon of a higher frequency passes through an optical material and splits into two photons of lower frequencies\cite{klyshko2018}. The photon at the higher frequency is referred to as the pump photon and the two photons at lower frequencies are referred to as the signal and the idler photons. 
SPDC is called parametric since the total energy remains conserved in the process. The constraints of energy and momentum conservations in SPDC require that the sum of the energies of the signal and idler photons be equal to the energy of the pump photon and that the sum of the momenta of the signal and idler photons be equal to the momentum of the pump photon. The SPDC process was first predicted theoretically by Louisell \etal in 1961 \cite{louisell1961pr}, who modelled  it as two oscillators coupled via a time-dependent reactance. Later, this effect was studied by many others before its first experimental observation as parametric fluorescence in a  $\rm LiNbO_3$ crystal \cite{harrish1967prl}. In this experiment, an intense argon-ion laser beam was used as a pump and the parametric fluorescence was observed over a significant portion of the visible and near-infrared spectrum through temperature tuning of the crystal. A similar effect was observed in an experiment with ammonium dihydrogen phosphate (ADP) crystal. In this case the spectral tuning was achieved by rotating the ADP crystal around the pump-beam
direction \cite{magde1967prl}. Later, instead of treating SPDC as a coupled oscillator problem,  Giallorenzi \etal\cite{giallorenzi1968pr} studied it as a scattering problem in which a pump photon transforms into a pair of photons due to nonlinear polarizability. In their study,  Giallorenzi \etal\cite{giallorenzi1968pr} were able to incorporate the effects due to factors such as the anisotropy and dispersion of the nonlinear crystal, input beam of finite spectral width, etc. Around the same time, D.N. Klyshko and co-authors studied SPDC as part of their investigations of spontaneous scattering due to quadratic and cubic terms in the expansion of the polarizability  \cite{klyshko1969jetplett, klyshko1969jetp}. They obtained formulas for the intensity of scattered light and  its dependence on the frequency and observation direction. Following up on the work by Klyshko and co-authors, Burnham \etal\cite{burnham1970prl} experimentally studied coincidences between the down-converted signal and idler photons and thereby demonstrated for the first time   that in the SPDC process the two photons are emitted simultaneously. This pioneering coincidence-detection experiment essentially marks the beginning of using SPDC photons for studying foundations of quantum mechanics and for exploring quantum information-based applications.

In nonlinear optical processes, the constraints due to the conservation laws are referred to as the phase-matching conditions. In the SPDC process, the phase matching is affected by various crystal as well as pump parameters and decides the efficiency with which a pump photon gets down-converted and the emission directions into which the signal and idler photons get generated. The SPDC phase-matching has been extensively studied in the past. It was shown by Franken \etal\cite{franken1961prl} that in order to observe the second-order nonlinear effect of spontaneous parametric down-conversion, one not only requires a highly intense pump field  but also a non-centrosymmetric crystal. In the context of second-harmonic generation in KDP crystals, Giordmaine \cite{giordmine1962prl} and Maker\cite{maker1962prl} independently  showed that the birefringence and dispersive properties of the crystal can be utilized for achieving the required phase matching conditions. Midwinter \etal \cite{midwinter1965britj} extended this technique to three-wave interaction processes, such as SPDC, for achieving the conditions for phase matching. Using the Kleinman's symmetry conjecture, Midwinter \etal \cite{midwinter1965britj} showed that the different types of phase-matching are satisfied by different crystals with appropriate symmetry properties \cite{kleinman1962pr}. The phase matching conditions in SPDC also decide the polarizations that the pump, the signal and the idler photons can have. For a negative uniaxial birefringent crystal, such as BBO\cite{midwinter1965britj,nikogosyan1991apa,yariv1988,nye1985, zernike2006}, in which the refractive index of light propagating along the optic axis of the crystal is less than the refractive index of light propagating perpendicular to it, there are a total of eight possible combinations of polarization states that the pump, signal and idler photons can take. However, due to the  phase matching constraints, only a few combinations are allowed. In general, when both the signal and idler photons have the same polarization, the phase-matching is referred to as the type-I phase matching. On the other hand, when the signal and idler photons have orthogonal polarizations, it is referred to as the type-II phase matching. Furthermore, if the directions of propagation of the down-converted signal and idler photons are along the propagation direction of the pump photon then it is known as the collinear phase-matching. Alternatively, in non-collinear phase matching, the signal and idler photons propagate in directions non-collinear with that of the pump photon.

The two SPDC photons have been shown to be entangled in various degrees of freedom such as energy-time \cite{kwiat1993pra,strekalov1996pra,khan2006pra,khan2007prl}, position-momentum \cite{law2004prl,howel2004prl,walborn2007pra}, polarization \cite{rubin1994pra,kwiat1994pra,walborn2003pra}, and angle-angular momentum \cite{rarity1990prl,mair2001nat,leach2010science}. In the last few decades, entanglement of the SPDC photons has been used for probing the foundations of quantum mechanics. One major step in this direction has been the ruling out of any local hidden variable interpretations of quantum mechanics \cite{bohm1952pr, bohm1952pr2} through the experimental violations of Bell's inequality \cite{bell1964physics, clauser1969prl}. Such violations have been observed in various degrees of freedom
including polarization \cite{ou1988prl, shih1988prl}, position and
momentum \cite{rarity1990prl}, time and energy
\cite{franson1989prl, kwiat1993pra, brendel1991prl},
spatial-parity \cite{yarnall2007prl}, and OAM \cite{leach2009optexp}. Using hyperentangled states of the SPDC photons, even simultaneous violations of Bell's
inequalities in more than one degree of freedom have been
reported \cite{strekalov1996pra, cinelli2005prl, yang2005prl,
barriero2005prl}. In addition to being used for probing the foundations of quantum mechanics, SPDC photons have also been used for exploring the entanglement-based real-world  applications such as quantum information\cite{bouwmeester1997nature,pittman2002pra,walther2005nat,tittel2001qinfcomp,irvine2004prl}, quantum cryptography\cite{tittle2000prl,jennewein2000prl,sergienko1999pra}, quantum dense coding\cite{mattle1996prl,bouwmeester1997nature}, optical measurements \cite{sergienko1995josab,dauler1999}, imaging \cite{pittman1995pra,dangelo2004prl,strekalov1995prl}, spectroscopy \cite{kira2011natphy}, quantum lithography and metrology \cite{boto2000prl,dangelo2001prl,matthews2016npj}.

SPDC has been experimentally realized using a variety of bulk crystals such as lithium niobate $(\rm LiNbO_3)$ \cite{harrish1967prl,klishko1968jetp}, $~\beta$-barium borate (BBO)\cite{kwiat1995prl,vanherzeele1988ao,cheng1988apl}, potassium dihydrogen phosphate (KDP)\cite{akhmanov1967jetplett}, ammonium dihydrogen phosphate (ADP)\cite{magde1967prl,giollorenzi1968apl}, and $\rm Ba_2NaNb_5O_{15}$\cite{krunger1970jap,byer1969jap} crystals. The bulk crystals such as BBO do not have very high down-conversion efficiency. Therefore, in more recent years, periodically-poled, quasi-phase matched waveguide crystals are being used to achieve higher SPDC efficiency, although in a restricted phase-matching range. Several experiments have reported SPDC generation using periodically-poled potassium titanyl phosphate (PPKTP) waveguide crystals \cite{fiorentino2007optexp,trojek2008apl,steinlechner2014josab}. The other quasi-phase matched waveguide crystals that are being used for SPDC generation include the periodically-poled lithium niobate $\rm(PPLN)$ crystals \cite{tanzilli2001ieee,martin2010njp}, and the periodically-poled lithium tantalate (PPLT) crystals \cite{leng2011natcom}.

In the past, although most of the foundational experiments \cite{kwiat1993pra,strekalov1996pra,khan2006pra,khan2007prl,law2004prl,howel2004prl,walborn2007pra, rubin1994pra,kwiat1994pra,walborn2003pra, rarity1990prl,mair2001nat,leach2010science, ou1988prl, shih1988prl, franson1989prl, brendel1991prl, yarnall2007prl, leach2009optexp, cinelli2005prl, yang2005prl,
barriero2005prl} as well as experiments exploring applications of photonic entanglement \cite{bouwmeester1997nature,pittman2002pra,walther2005nat,tittel2001qinfcomp,irvine2004prl, tittle2000prl,jennewein2000prl,sergienko1999pra, mattle1996prl, sergienko1995josab,dauler1999, pittman1995pra,dangelo2004prl,strekalov1995prl, kira2011natphy, boto2000prl,dangelo2001prl,matthews2016npj} have used BBO crystals for producing entangled photons by SPDC, the more recent experimental studies that require high down-conversion efficiencies in limited emission directions have started using the quasi-phase matched periodically-poled crystals for the purpose. Nevertheless, the current experimental studies, either foundational or application-oriented, that do not necessarily require very high down-conversion efficiencies still employ BBO crystals. This is due to the fact that the quasi-phase matching process that enables the periodically poled crystals to have high-down-conversion efficiencies compared to the bulk crystals does so at the expense of limiting the phase-matching ranges. As a consequence, although the periodically poled crystals become highly efficient for collinear down-conversion, the quasi-phase matching in such crystals does not allow for non-collinear down-conversion. Therefore, while these crystals can be used for studying temporal and polarization correlations of entangled photons, they are not effective in studying the spatial correlations of entangled photons in non-collinear emission geometries. Furthermore, the phase matching in the periodically-poled crystals is achieved by changing the temperature of the crystals, and therefore these crystals require temperature tuning and stability for achieving the desired phase matching condition. On the other hand, in BBO crystals, broad phase matching ranges can be very easily achieved. Both the collinear and non-collinear phase matching requirements can be satisfied simply by changing the angle between the optic axis of the crystal and the pump propagation direction. Moreover, these crystals have broad transmission ranges, have high damage threshold, and are readily available with various thicknesses. Therefore, a substantial number of experiments, especially those investigating spatial correlations of entangled photons, still employ BBO crystals for generating entangled photons via SPDC.

Because of the wide use of BBO crystals in generating SPDC photons with a variety of phase-matching geometries, several works in the past have investigated different aspects of SPDC phase matching in BBO crystals \cite{rubin1994pra, boeuf2000opteng, monken1998pra, kurtsiefer2001jmo, molina2005pra, schneeloch2016jo, walborn2010pr, rubin1996pra, pittman1996pra, klyshko2011physical}. However, a comprehensive study of SPDC phase-matching in BBO crystals covering all the major aspects of phase matching worked out in one place is still lacking. Such a study would be very useful to researchers working in various capacities with entangled  photons produced by SPDC in BBO crystals. In this tutorial article, we present a thorough study of the generation of entangled two-photon fields by the SPDC process in BBO crystal. We discuss the SPDC phase matching in various details and study how it is affected by various crystal and pump parameters. Through our simulation and experimental results, we also highlight several facts that would be very useful during experimental investigations with SPDC photons. Most of the calculations presented in this article have been carried out with the wavefunction of the down-converted photons expressed in the position or momentum bases. However, in the last section of this article, we discuss the two-photon wavefunction in the orbital angular momentum (OAM) basis and show how to calculate the OAM spectrum of the entangled two-photon field.

\subsection{Outline of the paper}\label{outline}

This article is organized as follows. In Sec.~\ref{spdc}, we discuss the spontaneous parametric down-conversion process in  $\beta$-barium borate (BBO) crystal. First, we derive the SPDC Hamiltonian and thereby work out the two-photon state produced by SPDC. We study how different pump and crystal parameters affect the phase matching and thus the generation of the signal and idler photons in SPDC. Our study includes effects due to several experimental factors such as crystal length, angle between optic axis and beam propagation direction, pump incidence angle on the crystal surface, refraction at the crystal surface, pump propagation inside the crystal, and the location of the beam waist of a Gaussian pump field. All these effects are extremely relevant in experimental situations. In Sec.~\ref{phase_matching}, we discuss different polarization effects in SPDC. We investigate how phase matching conditions decide the possible combinations of polarizations that the signal, idler and the pump photons can have, and how the polarization of these photons affect their propagation inside the nonlinear medium.  In Sec.~\ref{det-prob}, we calculate the detection probabilities of the down-converted photons for both type-I and type-II phase matching conditions. Furthermore, we calculate the conditional detection probability in the position basis. We then present various experimental and numerical results in  Sec.~\ref{result} in order to highlight how various experimental parameters affect the phase matching and thus the generation of entangled photons. In this context, we specifically point out that the intensity of the down-converted photons remains maximum for the collinear phase-matching and decreases from the maximum value with increasing non-collinearity. Nevertheless, the total power of the generated photons stays independent of the non-collinearity of phase matching. In Sec.~\ref{angular_spectrum}, we derive the two-photon wavefunction in the OAM basis and present the theoretical results that are useful for studying entanglement in the OAM basis. Finally, we present the conclusions in Sec.~\ref{conclusion}.

We would like to point out that the general formalism of Section 2 involving the derivation of the interaction Hamiltonian for SPDC and thereby the two-photon wavefunction is applicable to SPDC process not only in BBO crystals but also in any medium that has second-order nonlinearity. Furthermore, the general formalism presented in Section 4 for calculating the coincidence count rate, individual photon count rates, and the conditional detection probabilities is also applicable to SPDC process in any nonlinear medium and not just in BBO crystals. So, while the present tutorial article presents a complete study of phase-matching and SPDC-photon generation in BBO crystals, the general formalism presented in this article can be used for studying the SPDC phase-matching and thereby the entangled photons generation in any other nonlinear medium. Therefore, we expect this tutorial article to be useful in general to researchers working in various capacities with entangled photons produced via SPDC.

\section{Spontaneous parametric down-conversion}\label{spdc}

\subsection{Electromagnetic field inside a nonlinear medium}\label{em_wave}

Interaction of classical electromagnetic waves with a material medium is described by the Maxwell's equations. These equations are first-order coupled differential equations. The solutions to these equations provide information about the behavior of electric and magnetic fields inside a medium. The Maxwell's equations in a medium devoid of free charges and currents are given as 
\begin{eqnarray}
{\bm \nabla }\cdot {\bm D}({\bm r},t)  =0, \\
{\bm \nabla }\cdot {\bm B}({\bm r},t) =0,\\
{\bm \nabla }\times {\bm E}({\bm r},t) =- \frac{\partial{\bm B}({\bm r},t) }{\partial t} ,\label{eq:1}\\
{\bm \nabla }\times {\bm H}({\bm r},t) =  \frac{\partial{\bm D}({\bm r},t) }{\partial t},\label{eq:2}
\end{eqnarray}
where ${\bm E}({\bm r},t)$ stands for the electric field, ${\bm B}({\bm r},t)$ stands for the magnetic flux density, ${\bm H}({\bm r},t)=\frac{1}{\mu}{\bm B}({\bm r},t)$ is termed as magnetic field, where $\mu$ is the magnetic permeability. One can write $\mu=\mu_r \mu_0 $, where $\mu_r$ is the relative permeability of the medium and $\mu_0$ is called the magnetic permeability of free space. ${\bm D}({\bm r},t)$ stands for the electric displacement vector. Here, the medium is assumed to be magnetically isotropic, with the value of $\mu_r$ equal to one, but electrically anisotropic. The electric displacement vector ${\bm D}({\bm r},t)$ inside the medium is given by \cite{boyd2003,bloembergen1980josa} 
\begin{equation} \label{eq:3}
{\bm D}({\bm r},t) = \epsilon_0 {\bm E }({\bm r},t) + {\bm P }({\bm r},t),
\end{equation}
where ${\bm P }({\bm r},t)$ is termed as polarization. When the pump field strength is weak, the polarization is given by  ${\bm P }({\bm r},t)= \epsilon_0 \chi^{(1)}{\bm E }({\bm r},t)$, where $\chi^{(1)}$ is the linear susceptibility. However, when the pump field strength is strong, the polarization has higher-order contributions given by
\begin{eqnarray} \label{eq:4}
{\bm P}({\bm r},t) &= \epsilon_0\chi^{(1)} {\bm E }({\bm r},t) + \epsilon_0\chi^{(2)}{\bm E }({\bm r},t) {\bm E }({\bm r},t) \nonumber \\
& \qquad \qquad + \epsilon_0\chi^{(3)}{\bm E }({\bm r},t) {\bm E }({\bm r},t){\bm E }({\bm r},t)+\cdots, \nonumber\\
&={\bm P}^{(1)}({\bm r},t)+{\bm P}^{(2)}({\bm r},t)+{\bm P}^{(3)}({\bm r},t)+\cdots,
\end{eqnarray}
where $\chi^{(2)}$ is called the second-order nonlinear susceptibility, $\chi^{(3)}$ is called the third-order nonlinear susceptibility, etc. ${\bm P}^{(1)}({\bm r},t)$ is called the linear polarization while ${\bm P}^{(2)}({\bm r},t)$ is called the second-order nonlinear polarization, etc.  Typically, $\chi^{(2)}$ and $\chi^{(3)}$ are several orders of magnitude smaller than $\chi^{(1)}$ and as a result we see nonlinear effects only at very high field strengths. Using equation (\ref{eq:1}) through (\ref{eq:4}), we write the wave equation inside a nonlinear medium as
\begin{eqnarray} \label{eq:5}
{\bm \nabla }\left({\bm \nabla }\cdot{\bm E }({\bm r},t)\right) - { \nabla }^2{\bm E }({\bm r},t)= \nonumber \\
 - \mu_0\frac{\partial^2 }{\partial t^2}\left[\epsilon_0\left(1 + \chi^{(1)}\right){\bm E }({\bm r},t)  +{\bm P}^{(2)}({\bm r},t)+\cdots \right].
\end{eqnarray} 
Here $1 + \chi^{(1)} = n^2$ with $n$ being the refractive index of the medium. The susceptibilities above are tensor quantities, and for the second-order nonlinear effects we explicitly write the second-order nonlinear polarization as 
\begin{equation}
({\bm P}^{(2)})_l=\epsilon_0 \chi^{(2)}_{lmn}({\bm E })_m ({\bm E })_n. \label{Pi}
\end{equation}
Here, $l$, $m$, and $n$ are cartesian indices and $\chi^{(2)}_{lmn}$ is the second-order susceptibility tensor  \cite{ fowles1975 , boyd2003}. In this article, we mostly discuss the second-order nonlinear optical effect of spontaneous parametric down-conversion. 


\subsection{Introduction to SPDC}

\begin{figure}[t!]
\centering
\includegraphics[scale=0.8]{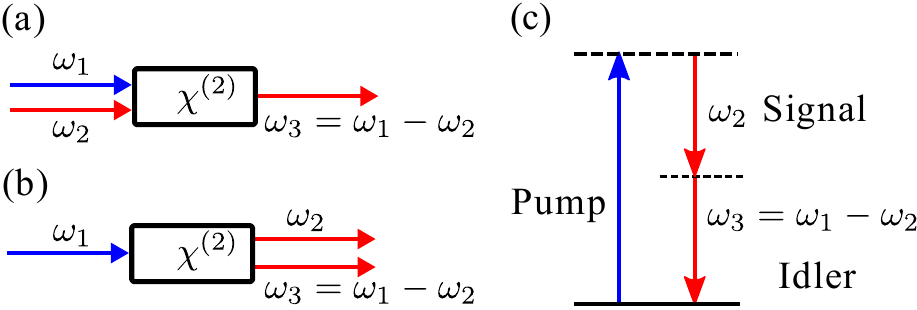}
\caption{(a) Difference frequency generation process (b) Spontaneous parametric down conversion process (c) Energy-level diagram of SPDC.}
\label{energy diagram}
\end{figure}

There are a variety of second-order nonlinear effects that take place inside a nonlinear medium, such as second-harmonic generation, sum-frequency generation, and difference-frequency generation. However, the second-order nonlinear effect that we concentrate on in this article is the spontaneous parametric down-conversion (SPDC). In SPDC, a single input photon at higher frequency, called the pump photon, interacts with a nonlinear crystal and splits into two photons of lower frequencies, called the signal and idler photons \cite{burnham1970prl,hong1985pra}. The SPDC is a type of difference-frequency generation process and is depicted in figure~\ref{energy diagram}(a). We note that in the process of difference-frequency generation, if the field at frequency $\omega_2$ is present in the input field, it stimulates the difference frequency generation at $\omega_3=\omega_1-\omega_2$, and as a result the process is usually very efficient and is called optical parametric amplification/oscillation. Even when the frequency $\omega_2$ is not present in the input field, one can still have the generation of difference frequency at $\omega_3=\omega_1-\omega_2$ accompanied by the generation of frequency $\omega_2$. However, in this case the generation of the difference frequency is stimulated by the presence of the vacuum mode at frequency $\omega_2$ in the input field and is therefore very inefficient. This second-order process of generating the frequencies $\omega_3$ and $\omega_2$ from the input frequency of $\omega_1$ is referred to as the spontaneous parametric down-conversion [see figure~\ref{energy diagram} (b)] . The term spontaneous refers to the fact that the process is stimulated by the vacuum mode at frequency $\omega_2$ in the input field and the term parametric signifies that the total energy remains conserved during the process.  Figure~\ref{energy diagram}(c) represents the energy-level diagram of the SPDC process, in which the photon at frequency $\omega_1$ gets absorbed and the absorber goes to a virtual state. From there, it decays down to the ground state emitting photons at frequencies $\omega_2$ and $\omega_3=\omega_1-\omega_2$. The constraints due to energy and momentum conservation make the two down-converted photons entangled in various degrees of freedom including energy-time \cite{strekalov1996pra,khan2006pra}, position-momentum \cite{law2004prl,walborn2007pra}, polarization \cite{walborn2003pra , kwiat1994pra} , and angular position-orbital angular momentum \cite{mair2001nat,rarity1990prl}. These entangled photons generated by SPDC have now become very important not only for studying the foundations of quantum mechanics but also for real-world applications of quantum entanglement.

\subsection{Hamiltonian for SPDC}

In order to study SPDC in all the details, we first need to derive the Hamiltonian for the process. For this, we start with the total (electrical plus magnetic) energy density of the
electromagnetic field, which is given by
\begin{eqnarray}
\bm D(\bm r,t)\cdot\bm E(\bm r,t)= \left[\epsilon_{0} \bm E(\bm r,t) +\epsilon_{0}\chi^{(1)}\bm E(\bm r,t)\right]\cdot \bm E(\bm r,t)  \nonumber\\ \qquad\qquad + \left[\epsilon_{0}\chi^{(2)}\bm E(\bm r,t) \bm E(\bm r,t)\right]\cdot \bm E(\bm r,t)+\cdots
\end{eqnarray}
The first term is the contribution to the electromagnetic energy
due to the linear term. The second term is the contribution due to
the second-order nonlinear term, etc. Since we are interested in
second-order nonlinear optical effects, we will look at the
second-order contribution to the energy, which is given by
\begin{eqnarray}
H_I(t) &=\int_{\mathcal V} \bm D(\bm r,t)\cdot\bm E(\bm r,t) d^3\bm r, \nonumber \\
& = \epsilon_0 \int_{\mathcal{V}}  \chi^{(2)} \bm E(\bm r,t) \bm E(\bm r,t) \cdot \bm E(\bm r,t)d^3\bm r,\nonumber\\
&=\int_{\mathcal V}  \bm P^{(2)}({\bm r},t)\cdot \bm E(\bm r,t)d^3\bm r,
\end{eqnarray}
where the integration extends over the volume $\mathcal{V}$ of the
nonlinear medium. This is the general expression for the
contribution to the total energy due to second-order nonlinear
optical effects. However, we are interested only in the nonlinear optical process of parametric down-conversion. So, using equation (\ref{Pi}), we write $H_I(t)$ as \cite{kleinman1968pr,ou1989pra}
\begin{eqnarray}
  H_I(t)= \epsilon_0\int_{\mathcal{V}} \chi_{lmn}^{(2)} \nonumber\\
 \qquad\qquad \times(\bm E_p(\bm r,t))_l(\bm E_s(\bm r,t))_m(\bm E_i(\bm r,t))_n d^3\bm r, \label{pdc-hamiltonian}
\end{eqnarray}
where $p$, $s$, and $i$ stand for pump, signal, and idler, respectively.In writing the above expression, we have used the vectorial form of the electric fields. This is because of the fact that the down-conversion process does depend on the polarizations of the pump, signal and idler fields. However, working with vectorial fields  makes the calculations quite cumbersome. Therefore, our approach in this article is to first find out the allowed combinations of polarization states that the signal, idler and pump photons can have and then for each combination, work with the scalar fields as far as calculating the detection probabilities of the down-converted photons are concerned. To this end, we write $H_I(t)$ as
\begin{eqnarray}
  H_I(t)= \epsilon_0\int_{\mathcal{V}} \chi^{(2)} E_p(\bm r,t)E_s(\bm r,t)E_i(\bm r,t) d^3\bm r, \label{pdc-hamiltonian2}
\end{eqnarray}
where $\chi^{(2)}$ stands for the second-order nonlinearlity, and $E_p(\bm r,t)$, $E_s(\bm r,t)$, and $E_i(\bm r,t)$ are the scalar electric fields for an allowed combination of polarizations of the signal, idler and pump photons. 

In quantum formalism, the contribution $H_I(t)$ to the energy  takes the form of the interaction Hamiltonian operator $\hat{H}_I(t)$ and can be written as 
\begin{eqnarray}
  \hat{H}_I(t)= \epsilon_0\int_{\mathcal{V}} \chi^{(2)} \hat{E}_p(\bm r,t)\hat{E}_s(\bm r,t)\hat{E}_i(\bm r,t) d^3\bm r. \label{pdc-hamiltonian3}
\end{eqnarray}
Here, $\hat{E}_p(\bm r,t)$, $\hat{E}_s(\bm r,t)$, and $\hat{E}_i(\bm r,t)$ are the electric field operators corresponding to the pump, signal and idler fields, respectively. Next, we write the fields in terms of their negative and positive frequency parts. This procedure is equivalent to the complex analytic signal representation for classical fields \cite{mandel1967josa}. Although the complex analytic signal representation was introduced as an outstanding tool for simplifying the mathematical handling of classical fields, its counterpart in the context of quantum field has become a necessary ingredient for handling creation and annihilation of photons. Following Ref. \cite{glauber1963pra}, we write a quasi-monochromatic quantum electric field operator $\hat{E}(t)$ as
\begin{eqnarray}\label{eq:6aa}
\hat{E}(t) &= \int_{-\infty}^{\infty}\hat{\tilde{E}}(\omega) e^{-i \omega t} d\omega, \nonumber \\  
 & = \int_{-\infty}^{0}\hat{\tilde{E}}(\omega) e^{-i \omega t}d\omega + \int_{0}^{\infty}\hat{\tilde{E}}(\omega) e^{-i \omega t} d \omega,\nonumber \\ 
 &= \hat{E}^{(-)}(t) +     \hat{E}^{(+)}(t) .  
\end{eqnarray}
Here $\hat{E}^{(+)}(t)$ and $\hat{E}^{(-)}(t)$ are the positive and negative frequency parts of the electric field operator, and since the electric field operator $\hat{E}(t)$ is Hermitian, we have $\hat{\tilde{E}}(\omega)=\hat{\tilde{E}}^{\dagger}(-\omega)$. Representing this way, we write the pump, signal and idler electric field operators in the following form
\begin{eqnarray*}
\hat{E}_p(\bm r, t) = \hat{E}^{(+)}_p (\bm r, t)+ \hat{E}^{(-)}_p(\bm r, t),\\
\hat{E}_s(\bm r, t) = \hat{E}^{(+)}_s(\bm r, t) + \hat{E}^{(-)}_s(\bm r, t),\\
\hat{E}_i(\bm r, t) = \hat{E}^{(+)}_i (\bm r, t)+ \hat{E}^{(-)}_i(\bm r, t).
\end{eqnarray*}
Here $\hat{E}^{(+)}_p (\bm r, t)$ is the positive frequency part of the pump electric field operator, etc. Using the above forms for the field operators, the Hamiltonian in equation (\ref{pdc-hamiltonian3}) can be written as
\begin{eqnarray} \label{int_Hamiltonian}
 \hat{H}_I(t) = \epsilon_0\int_{\mathcal{V}}\chi^{(2)}\left[\hat{E}^{(+)}_p({ \bm r },t) + \hat{E}^{(-)}_p({ \bm r },t)\right] \nonumber \\  \times\left[\hat{E}^{(+)}_s({ \bm r },t) + \hat{E}^{(-)}_s({ \bm r },t)\right]\left[ \hat{E}^{(+)}_i({ \bm r },t) + \hat{E}^{(-)}_i({ \bm r },t)\right] d^3\bm r. \nonumber
\\  
\end{eqnarray}
The resulting expression for the interaction Hamiltonian is a sum of eight different terms with all possible
combinations of the three fields. However, there are only two
terms, $\hat{E}_p^{(+)}({ \bm r },t)\hat{E}_s^{(-)}({ \bm r },t)\hat{E}_i^{(-)}({ \bm r },t)$ and $\hat{E}_p^{(-)}({ \bm r },t)\hat{E}_s^{(+)}({ \bm r },t)\hat{E}_i^{(+)}({ \bm r },t)$, that lead to energy
conserving processes and thus contribute appreciably to the
down-conversion process. The contributions due to the other six
terms get averaged out when the
interaction Hamiltonian $\hat{H}_I(t)$ is integrated over time.
Therefore, we neglect the contributions due to these other terms. Neglecting these contributions is equivalent to making the
rotating-wave approximation as in the case of treating atomic
absorption and emission processes (see Ref.~\cite{loudon2000}, Section
2.3). We note that these approximations hold only for second-order nonlinear optical
processes such as SPDC and that for the higher-order nonlinear optical processes the
non-energy-conserving terms may lead to important contributions.
The effective interaction Hamiltonian for the process of
parametric down-conversion can thus be given by the following
simplified form:
\begin{eqnarray} \label{eq:10}
 \hat{H}_I(t) =
  \epsilon_0\int_{\mathcal{V}} \chi^{(2)} \hat{E}_p^{(+)}({ \bm r },t) \nonumber \\
  \qquad\qquad \times \hat{E}_s^{(-)}({ \bm r },t) \hat{E}_i^{(-)}({ \bm r },t)d^3\bm r + \rm{H.c.} ,
\end{eqnarray} 
where $\rm H.c.$ stands for Hermitian conjugate. Next, we need to express the above Hamiltonian in terms of the photon creation and annihilation operators. For this purpose, we first realize that a classical electric field $E ({\bm r}, t)$ can be represented  in terms of its spectral decomposition as
\begin{eqnarray}
E ({\bm r}, t)= \int_{-\infty}^\infty\tilde{E}({\bm r},\omega)e^{-i\omega t} d\omega, \nonumber \\
\qquad = \int_{-\infty}^\infty \tilde{E}({\bm \rho},z;\omega)e^{-i \omega t} d\omega, \nonumber \\
\qquad= \int_{-\infty}^\infty\hspace{-1.5mm}\left( \int A~ a({\bm q}, \omega)e^{ i \left({\bm q \cdot \bm\rho} + k_z z \right)} d^2 {\bm q}\right) e^{-i \omega t} d\omega.
\end{eqnarray} 
Here A is a constant, $\bm k=(k_x, k_y, k_z)=(\bm q, k_z)$, $\bm r=(x, y, z)=(\bm\rho, z)$, and $a(\bm q, \omega)$ is the angular spectrum representation of the field. We know that the spectral decomposition of a field remains the same irrespective of whether it is represented within classical or quantum description. Therefore, we use the above representation to write the quantized electric fields simply by replacing $a({\bm q}, \omega)$ with the corresponding annihilation operator $\hat{a}({\bm q}, \omega)$ \cite{glauber1963pra,grynberg&aspect2010}. This way, we write the positive-frequency part of the signal, idler and pump field operators as
\begin{eqnarray} \label{eq:14}
\hat{E}_s^{(+)}({ \bm r },t) = \int\!\!\!\!\int d^2 {\bm q}_s d\omega_s A_s e^{ i \left({\bm q}_s \cdot {\bm \rho} +k_{sz} z - \omega_s t\right)}  \hat{a}({\bm q}_s, \omega_s), \nonumber \\
\hat{E}_i^{(+)}({ \bm r },t) = \int\!\!\!\!\int d^2 {\bm q}_i d\omega_i A_i e^{ i \left({\bm q}_i \cdot {\bm \rho} +k_{iz} z - \omega_i t\right)}  \hat{a}({\bm q}_i, \omega_i),  \nonumber \\
\hat{E}_p^{(+)}({ \bm r },t) = \int\!\!\!\!\int d^2 {\bm q}_p d\omega_p A_p e^{ i \left({\bm q}_p \cdot {\bm \rho} +k_{pz} z - \omega_p t\right)}  V({\bm q}_p, \omega_p).\nonumber \\
\end{eqnarray}
Here, we note that in quantizing the electromagnetic fields one ends up with a summation over wave-vectors. However, in the limit when the quantization volume goes to infinity, the discrete set of wave-vectors becomes continuous and the summation can be replaced by an integral. In writing the form of the field operators above, we have made use of this limiting condition. We further note that the intensity of the pump field is usually several orders of magnitude larger than that of the signal and idler fields. Therefore, we treat the pump field classically and replace the annihilation operator corresponding to the pump field by the field amplitude $V({\bm q}_p, \omega_p)$. With this assumption and using the form of the field operators in equation (\ref{eq:14}), we write equation (\ref{eq:10}) as 
\begin{eqnarray} \label{eq:15}
\hat{ H}_I(t)=  \epsilon_0 \chi^{(2)} A_p A^*_s A^*_i \int_{\mathcal{V}}d^3\bm r \int \!\!\!\!\int \!\!\!\!\int d \omega_p d \omega_s d \omega_i  \nonumber \\
\times\int \!\!\!\!\int \!\!\!\!\int  d^2 {\bm q}_p d^2 {\bm q}_s d^2 {\bm q}_i V\left({\omega_p,{\bm q}_p }\right)  \nonumber \\
\times \mbox{exp} \left[ i({\bm q}_p-{\bm q}_s -{\bm q}_i  )\cdot{\bm \rho}+i\left(k_{pz}-k_{sz}- k_{iz}\right)z \right]  \nonumber \\
  \times e^{i(\omega_s +\omega_i -\omega_p)t} \hat{ a}^{\dagger} ({\bm q}_s,  \omega_s)\hat{ a}^{\dagger} ({\bm q}_i, \omega_i) + \mbox{H.c}.
\end{eqnarray}
Here, although $\chi^{(2)}$ and $A_j$ with $j=p$, $s$, and $i$ are frequency dependent quantities, they vary very slowly within the frequency ranges of interest for most experimental situations. So, we have taken them outside of the integral. Equation~(\ref{eq:15}) is the interaction Hamiltonian for spontaneous parametric down-conversion.

\subsection{The generation of entangled two-photon field}

\begin{figure}[t!]
\centering
\includegraphics[scale=0.9]{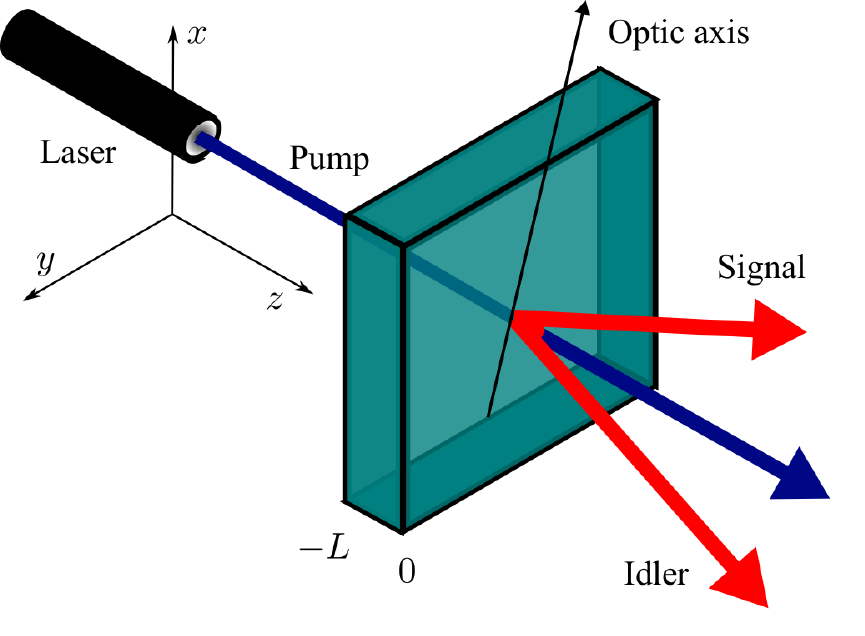}
\caption{A typical experimental arrangement for SPDC. The pump field is along the $z$ direction. The figure shows the laboratory frame coordinates of the nonlinear crystal of thickness $L$.  }
\label{fig:coor}
\end{figure}

We now derive the state of the entangled photons that gets generated in SPDC. Let us consider a pump field interacting with a nonlinear optical crystal of thickness $L$ (see Figure~\ref{fig:coor}). We assume that the pump photon starts interacting with the crystal at time $t=-t_0$ and that the state $\ket{\psi(-t_0)}$ of the down-converted signal and idler photons at that instant is given by $\ket{\psi(-t_0)}=\ket{\rm vac}_s\ket{\rm vac}_i$, where $|\rm vac\rangle$ represent a vacuum mode. The state of the two photons evolves under Hamiltonian $H_I(t)$ of equation (\ref{eq:15}) and at the end of the interaction, that is, at time $t=0$, the two-photon state is given by  
\begin{equation} \label{eq:11}
|\psi(0)\rangle = e^{-\frac{i}{\hbar}\int_{-t_0}^{0}\hat{H}_I(t)dt} | \psi(-t_0)\rangle.
\end{equation}
The parametric interaction is assumed to be very weak, so the
state in equation (\ref{eq:11}) can be approximated by
the first two terms of a perturbation expansion. The first term is
simply the initial vacuum state $\ket{\psi(-t_0)}$. The second term is the state of the  down-converted two-photon field and  is given by
\begin{eqnarray} \label{eq:12}
|\psi_{\rm tp}\rangle =  -\frac{i}{\hbar}\int_{-t_0}^{0}\hat{H}_I(t)dt \ket{\psi(-t_0)} 
\end{eqnarray}
The third and the higher-order terms in the expansion are the states of four- and higher-photon fields, and we assume that the probability of generation of these states are negligible. 
Substituting the expression of $\hat{H}_I(t)$ from equation (\ref{eq:15}) into the above equation, we obtain the state of the two-photon field at the exit face inside the nonlinear crystal:
\begin{eqnarray} \label{eq:17}
|\psi_{\rm tp}\rangle =\frac{\epsilon_0}{i\hbar}\chi^{(2)} A_p A^*_s A^*_i \int_{-t_0}^0 \!\!dt \int_{\mathcal{V}} d^2{\bm\rho}dz \int\!\!\!\!\int\!\!\!\!\int\!\!d\omega_p d\omega_s d\omega_i \nonumber \\  
\times \int \!\!\!\!\int \!\!\!\!\int d^2 {\bm q}_p d^2 {\bm q}_s d^2 {\bm q}_i V(\omega_p,{\bm q}_p) \nonumber\\
\times  \mbox{exp} \left[ i({\bm q}_p-{\bm q}_s -{\bm q}_i  )\cdot{\bm \rho} + i\left(k_{pz}- k_{sz}-k_{iz}\right)z \right]    \nonumber \\
\times \mbox{exp}\left\lbrace i(\omega_s +\omega_i -\omega_p)t\right\rbrace \nonumber \\ \times \hat{ a}^{\dagger} ( {\bm q}_s,  \omega_s)\hat{ a}^{\dagger} ({\bm q}_i, \omega_i) \ket{\rm vac}_s\ket{\rm vac}_i + \mbox{H.c.},
\end{eqnarray}
where we have substituted $d^3\bm r=\int\int d^2{\bm\rho}dz$.  The interaction time $t_0$ is assumed to be much longer than the time scale over which down-conversion takes place. Therefore, the limits of time integration can be extended to $-\infty$  and $\infty$ \cite{ou1989pra,grice1997pra} such that the time integration yields  
\begin{equation}\label{eq:18}
\int^{\infty}_{-\infty}e^{ i(\omega_s +\omega_i -\omega_p)t} dt = \delta\left(\omega_s +\omega_i - \omega_p\right).
\end{equation}
Similarly, we assume that the transverse area of the nonlinear crystal is much larger compared to the transverse area of the pump field, and therefore we write the space integral in equation (\ref{eq:17}) as
\begin{eqnarray}\label{eq:19}
\int^{\infty}_{-\infty}\mbox{exp}\left\lbrace i ({\bm q}_p -{\bm q}_s- {\bm q}_i)\cdot {\bm \rho} \right\rbrace  d^2{\bm \rho}  \nonumber\\
\qquad\times  \int^{0}_{-L}\mbox{exp}\left\lbrace i (k_{pz} -k_{sz}- k_{iz})z \right\rbrace  dz , \nonumber \\
= \delta\left({\bm q}_p -{\bm q}_{s}- {\bm q}_{i}\right) \Phi(\omega_s,\,\omega_i,{\bm q}_s,{\bm q}_i),  \\
{\rm where} \nonumber \\
\Phi(\omega_s,\,\omega_i,{\bm q}_s,{\bm q}_i)=\int^{0}_{-L}\mbox{exp}\left[ i (k_{pz} -k_{sz}- k_{iz})z \right]  dz , \nonumber \\
=L~\mbox{sinc}\left[ (k_{sz} +k_{iz}- k_{pz})\frac{L}{2}\right]\mbox{exp}\left[ i (k_{sz} +k_{iz}- k_{pz})\frac{L}{2}\right],\nonumber
\end{eqnarray}
is called the phase-matching function. The action of the creation operators in equation (\ref{eq:17}) is given by
\begin{equation}\label{creation}
\hat{ a}^{\dagger} ({\bm q}_s,  \omega_s)\hat{ a}^{\dagger} ({\bm q}_i, \omega_i) \ket{\rm vac}_s\ket{\rm vac}_i=\ket{ {\bm q}_s,\omega_s}_s \ket{ {\bm q}_i, \omega_i}_i, \label{creation}
\end{equation}
where $|{\bm q}_s, \omega_s\rangle_s$ represents a state having one signal photon with transverse wave-vector ${\bm q}_s$ and frequency $\omega_s$. The Hermitian conjugate part of $|\psi_{\rm tp}\rangle$ in equation (\ref{eq:17}) contains the operator $\hat{ a}({\bm q}_s,\omega_s)\hat{ a}({\bm q}_i,\omega_i)$. The action of this operator on the vacuum state is given by
\begin{equation}
\hat{ a}({\bm q}_s,\omega_s)\hat{ a}({\bm q}_i,\omega_i)|\mbox{vac}\rangle_s |\mbox{vac} \rangle_i = 0.\label{annihilation}
\end{equation}
Therefore, using equations (\ref{eq:18}) through (\ref{annihilation}), we write equation (\ref{eq:17}) as
\begin{eqnarray}\label{eq:23}
|\psi_{\rm tp}\rangle =A \int\!\!\!\! \int \!\!\!\!\int d \omega_p d \omega_s d \omega_i \int \!\!\!\!\int \!\!\!\!\int d^2 {\bm q}_p d^2 {\bm q}_s d^2 {\bm q}_i \nonumber \\ \qquad\quad \times V(\omega_p, {\bm q}_p)  \Phi(\omega_s,\,\omega_i,{\bm q}_s,{\bm q}_i) \delta\left(\omega_s +\omega_i - \omega_p\right)\nonumber \\ \qquad\quad \times \delta\left({\bm q}_s +{\bm q}_{i}- {\bm q}_{p}\right)\ket{ {\bm q}_s, \omega_s}_s \ket{ {\bm q}_i, \omega_i}_i ,
\end{eqnarray}
where $A=\frac{\epsilon_0}{i\hbar}\chi^{(2)} A_p A^*_s A^*_i$.
Now, integrating over $d\omega_p $ and $d {\bm q}_p$, we get
\begin{eqnarray}\label{psitp_detctor}
|\psi_{\rm tp}\rangle =A \int \!\!\!\!\int d \omega_s d \omega_i \int \!\!\!\!\int d^2 {\bm q}_s d^2 {\bm q}_i V(\omega_s+\omega_i, {\bm q}_s+{\bm q}_i) \nonumber \\  \qquad\qquad\times\Phi(\omega_s,\,\omega_i,{\bm q}_s,{\bm q}_i)  \ket{ {\bm q}_s, \omega_s}_s \ket{{\bm q}_i,\omega_i}_i .
\end{eqnarray}
The above equation represents the state of the two-photon field at the exit face of the crystal produced by SPDC.

\subsection{Effects due to the beam waist location of a Gaussian pump beam}

\begin{figure}[t!]
\centering
\includegraphics{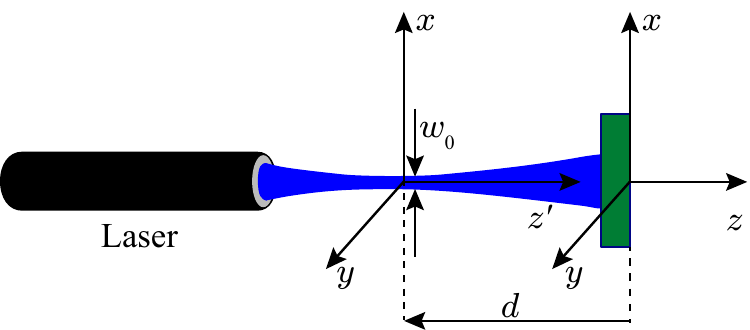}
\caption{Schematic diagram depicting the situation when the pump beam waist is located at a distance $d$ behind the crystal.  }
\label{beam-waist}
\end{figure}

In most situations, the pump field is taken to be in the form of a Gaussian beam and the location of the Gaussian beam waist is taken to coincide with that of the crystal. However, in many experimental situations, the location of the pump beam waist does not coincide with that of the crystal. Effects of this kind have been studied in several works in the past \cite{belinskii1994jetp, pittman1995pra,pittman1996pra,monken1998pra_lens,rubin1996pra}, and in this subsection, we present detailed calculation illustrating how the beam-waist location of the pump field affects the SPDC process. Let us consider the situation shown in figure~\ref{beam-waist}, in which the nonlinear crystal is at a distance $d$ away from the beam waist location of the Gaussian pump field. Our aim is to find out the form of the pump field $V({\bm q}_s + {\bm q}_i)$ that needs to be used in equation (\ref{coincidence}). In order to do this, we first define a coordinate system $(x,y,z')=(\bm\rho, z')$ such that the location of the pump beam waist is at $z'=0$. We define another coordinate system $(x,y,z)=(\bm\rho, z)$ such that the crystal is located at $z=0$. Thus we have $z' = z + d$. The electric field amplitude $\tilde{V}(\bm\rho, z')$ at a given $z'$ can be written as 
\begin{equation}
\tilde{V}(\bm\rho, z')=\int\!\!\!\!\int V(\bm q_p)e^{i(\bm q_p\cdot \bm\rho + k_{pz}z')}d^2 \bm q , \label{field-z}
\end{equation} 
where  $V(\bm q_p)$ is the angular spectrum of the field at $z'=0$ and where we have used $\bm{k_p}= (\bm q_p, k_{pz})$. Now using the fact $z'=z +d $, we rewrite equation~(\ref{field-z}) in the $(\bm\rho,z)$ coordinate system as
\begin{equation}
\tilde{V}(\bm\rho, z)=\int\!\!\!\!\int V(\bm q_p)e^{i k_{pz}d} e^{i(\bm q_p\cdot \bm\rho + k_{pz}z)}d^2 \bm q.
\label{field-z'}
\end{equation} 
Here $\tilde{V}(\bm\rho, z)$ is the same electric field amplitude  but written in the  $(\bm\rho,z)$ coordinate system. Comparing equations (\ref{field-z}) and (\ref{field-z'}), we see that $V(\bm q_p)e^{i k_{pz}d}$ can be taken as the angular spectrum of the field at $z=0$. Therefore, we find that if the angular spectrum of the pump field is known at a distance $d$ behind the crystal, the angular spectrum at the crystal plane can simply be taken as $V(\bm q_p)e^{i k_{pz}d}$. 

In most cases, the pump field satisfies the paraxial approximation. Under this approximation, we write $k_{pz}\approx k_p-|\bm q_p|^2/2k_p$, where $k_p=|\bm k_p|=\omega_{p0}/c=(\omega_{s0}+\omega_{i0})/c$. For a pump field with Gaussian spectrum, the angular spectrum at the  beam waist is given by \cite{milonni&eberly2010}
$ V(\bm q_p)=\exp\left[ -|\bm q_p|^2 w_0^2/4\right]$, where $w_0$ is the beam waist at $z=-d$. Therefore the angular spectrum of the field at the crystal location, that is, at $z=0$ can be written as
\begin{eqnarray}
\!\! V(\bm q_p)e^{i k_{pz}d}=\exp\left[ -\frac{|\bm q_p|^2 w_0^2}{4}\right]\exp\left[ -i\frac{|\bm q_p|^2 d}{2 k_p}\right]e^{ik_p d} .
\end{eqnarray}
We note that in the context of using the above spectrum for calculating the coincidence count rate in equation (\ref{coincidence}), the constant factor $e^{ik_p d}$ does not make any contribution. 
%
%
This is the expression for $V(\bm q_p)=V(\bm q_s+\bm q_i)$ that we will be using in equation (\ref{coincidence}) for calculating the coincidence count rates.

\subsection{Effects due to the non-normal incidence of pump field on the SPDC crystal}

\begin{figure}[t!]
\centering
\includegraphics[scale=0.9]{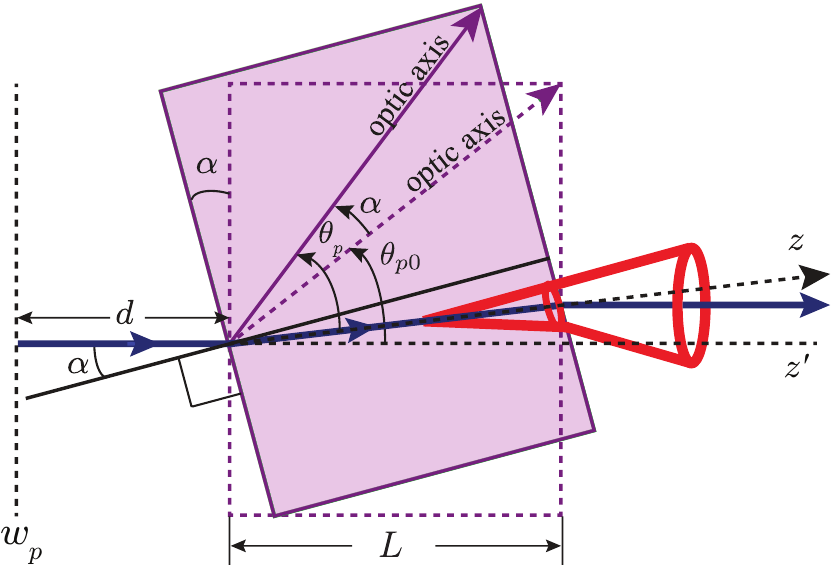}
\caption{Schematic diagram illustrating effects due to the non-normal incidence of the pump field on the nonlinear crystal.}
\label{internal_thetap}
\end{figure} 

In experimental situations, it is quite often the case that the SPDC crystal is oriented in such a way that the pump field is incident on the crystal at an angle other than $90^0$. In this subsection, we consider the effects on the two-photon coincidence count rate caused by the non-normal incidence. This situation is shown in figure \ref{internal_thetap}. The SPDC crystal is rotated by an angle $\alpha$ with respect to the pump propagation direction. Inside the crystal, the $z'$ axis represents the direction of propagation of the pump field when it is incident perpendicular to the crystal, that is, when $\alpha=0$, while the $z$ axis represents the direction of propagation of the pump beam inside the rotated crystal. The angle between the optic axis and the $z'$ axis is represented by $\theta_{p0}$ while $\theta_p$ represents the angle between the $z$ axis and the optic axis of the rotated crystal. Using Snell's law of refraction we write ${\rm sin} \alpha = n_{pe} {\rm sin}\beta$, where $\alpha$ is the angle of incidence of the pump field, $\beta$ is the  angle of refraction inside the crystal, and $n_{pe}$ is the refractive index of the pump field which is taken to be extra-ordinary polarized. Here, we note that the pump field polarization is decided by the phase-matching conditions, and we show in the next section through the phase-matching considerations that the pump field has to be extra-ordinary polarized for negative uniaxial crystals such as BBO. Therefore, we write
\begin{eqnarray}
\qquad \qquad \qquad \beta = {\rm sin}^{-1}\left(\frac{{\rm sin}~ \alpha}{n_{pe}}\right).
\end{eqnarray}
From the ray diagram shown in figure \ref{internal_thetap}, we obtain
\begin{eqnarray}
\qquad\qquad  \theta_p &= \theta_{p0} + \beta = \theta_{p0} + {\rm sin}^{-1}\left(\frac{{\rm sin}~\alpha}{n_{pe}}\right).
\end{eqnarray} 
Thus we see that the non-normal incidence of the pump field affects the angle between the optic axis and the pump propagation direction. Therefore, we need to use $\theta_p$ instead of $\theta_{p0}$ when calculating $k_{pz}$ and subsequently the phase-matching function $\Phi(\bm q_s, \bm q_i)$ in equation (\ref{psitp_detctor}).

\section{Polarization effects in SPDC}\label{phase_matching}

\subsection{Polarization effects due to phase matching constraints}

\begin{figure}[b!]
\centering
\includegraphics[scale=0.8]{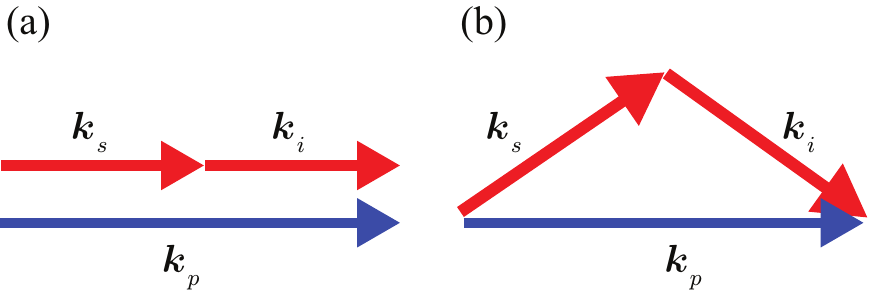}
\caption{Phase-matching diagrams for the (a) collinear and (b) non-collinear emission geometries. The condition for perfect phase-matching is $k_{pz}=k_{sz}+k_{iz}$.}
\label{fig:coll}
\end{figure}

The generation and emission-directions of SPDC photons are decided by the constraints imposed by the conservation of energy and momentum. These constraints are referred to as the phase-matching conditions. Numerous textbooks discuss phase matching based on  crystal class and permutation symmetries\cite{new&geoffrey2011,yariv&yeh1984,hutchings1992oqe,peter&joseph2017,nye1985, zernike2006}. If the directions of propagation of the down-converted signal and idler photons are along the propagation direction of the pump photon then it is known as the collinear phase matching. Alternatively, in non-collinear phase matching, the signal and idler photons propagate in directions non-collinear with that of the pump photon. The phase matching diagrams of collinear and non-collinear SPDC are shown in figure \ref{fig:coll}$\rm (a)$ and \ref{fig:coll}$\rm (b)$, respectively. As we see from equation (\ref{psitp_detctor}), the efficiency of the SPDC process, that is, the generation probablity of the down-converted photons depends on the phase-matching function $\Phi(\bm q_s, \bm q_i)$ through the wave vector mismatch $\Delta k_{z} = k_{sz}+  k_{iz}- k_{pz}$. It can be shown that this efficiency goes as $\mbox{sinc}^2\left[\Delta k_{z}L/2\right]$. When $\Delta k_{z}=0$, that is, when the phase matching is perfect, the efficiency is maximum. The efficiency of the SPDC process decreases as the phase mismatch $\Delta k_{z}$ increases \cite{maker1962prl}.

Next, we recall that in deriving the Hamiltonian in  equation (\ref{eq:15}), we had considered only the scalar fields even though polarization plays an important role in deciding the phase-matching condition. In this section, we consider the polarization effects due to the phase matching constraints. Our approach is to first find out the possible polarization scenarios that can satisfy the phase-matching condition and for those scenarios use equation (\ref{psitp_detctor}) for calculating the detection probabilities. In order to list the possible polarization scenarios that can satisfy the phase matching, we analyze it in the collinear configuration. In this configuration, the phase matching constraint $\Delta k_{z}=0$ requires the following equation to be satisfied (see Ref. \cite{boyd2003}, Section 2.3):
\begin{equation}\label{conserve_eq}
\frac{n_{p\sigma}(\omega_{p0}) \omega_{p0}}{c} = \frac{n_{s\sigma}(\omega_{s0}) \omega_{s0}}{c} + \frac{n_{i\sigma}(\omega_{i0}) \omega_{i0}}{c},
\end{equation}
where as earlier, $p$, $s$, and $i$ stand for pump, signal, and idler, respectively. The index $\sigma$ can take two values. If the polarization direction is perpendicular to the plane formed by the optic axis and the propagation direction then it is called ordinary polarized, and in that case $\sigma=o$. Alternatively, if the polarization direction is parallel to the plane formed by the optic axis and the propagation direction then it is called extraordinary polarized, and in that case $\sigma=e$ \cite{peter&joseph2017}. Therefore, $n_{pe}(\omega_{p0})$ represents the refractive index of the extra-ordinary polarized pump photon at frequency $\omega_{p0}$, etc. Combining the above equation with energy conservation equation, that is, $\omega_{p0} = \omega_{s0} +\omega_{i0}$, we rewrite equation~(\ref{conserve_eq}) as 
\begin{equation}\label{n_eq}
n_{p\sigma}(\omega_{p0})  - n_{s\sigma} (\omega_{s0})= \frac{\omega_{i0}}{\omega_{p0}}\left[n_{i\sigma}(\omega_{i0})-n_{s\sigma}(\omega_{s0})\right].
\end{equation}
Here, we note that $\omega_{p0} > \omega_{s0} \geq \omega_{i0}$. We now consider all possible polarization scenarios for pump, signal and idler photons one by one and find out the polarization scenarios that can potentially satisfy the phase matching condition. We assume normal dispersion for the down-conversion crystal, that is, we assume that the refractive index increases with frequency. BBO crystal is an example of a down-conversion crystal that has normal dispersion for both ordinary and extraordinary polarization \cite{eimerl1987jap}. There can be a total of 8 possible polarization scenarios for the pump, signal and idler photons. For each scenario, we represent the down-conversion process by a rightarrow and the individual pump, signal and idler photons only by their polarization index. For example, $ e \rightarrow o + e $ represents an extraordinary polarized pump down-converting into an ordinary signal photon and an extraordinary idler photon, etc.    
\linebreak
\\
{\bf Case I: $\bm{ e \rightarrow e + e :}$}\\
In this case, the pump, signal and idler photons are all extraordinary polarized and therefore equation (\ref{n_eq}) can be written as 
\begin{eqnarray}
n_{pe}(\omega_{p0})  - n_{se} (\omega_{s0})= \frac{\omega_{i0}}{\omega_{p0}}\left[n_{ie}(\omega_{i0})-n_{se}(\omega_{s0})\right]. 
\end{eqnarray}
We note that $n_{pe}(\omega_{p0}) >n_{se}(\omega_{s0}), n_{ie}(\omega_{i0})$. Considering $\omega_{s0} \geq \omega_{i0}$ we find that the above equation cannot be satisfied. Therefore, the phase matching condition can never be satisfied and thus this particular polarization scenario is not possible.
\linebreak
\\
{\bf Case II: $\bm{ o \rightarrow o + o :}$}\\
In this case, the pump, signal and idler photons are all ordinary polarized and therefore the equation (\ref{n_eq}) can be written as 
\begin{eqnarray}
n_{po}(\omega_{p0})  - n_{so} (\omega_{s0})= \frac{\omega_{i0}}{\omega_{p0}}\left[n_{io}(\omega_{i0})-n_{so}(\omega_{s0})\right]. 
\end{eqnarray}
We note that $n_{po}(\omega_{p0}) >n_{so}(\omega_{s0}), n_{io}(\omega_{i0})$. As we assume $\omega_{s0} \geq \omega_{i0}$, we find that the equation above cannot be satisfied. Therefore, the phase matching condition can never be satisfied and thus this particular polarization scenario is not possible.
\linebreak
\\
{\bf Case III: $\bm{ e \rightarrow o + o :}$}\\
In this case, equation (\ref{n_eq}) can be written as 
\begin{eqnarray}
n_{pe}(\omega_p)  - n_{so} (\omega_{s0})= \frac{\omega_{i0}}{\omega_{p0}}\left[n_{io}(\omega_{i0})-n_{so}(\omega_{s0})\right].
\end{eqnarray}
We find that for $\omega_{s0} \geq \omega_{i0}$, the right hand side of the above equation is negative. Now in order for the left hand side of this equation to be negative, we have to ensure that $n_o > n_e$. Therefore, this scenario can indeed be satisfied by negative uniaxial crystals such as BBO. 
\linebreak
\\
{\bf Case IV: $\bm{ o \rightarrow e + e :}$}\\
In this case, equation (\ref{n_eq}) can be written as 
\begin{eqnarray}
n_{po}(\omega_{p0})  - n_{se} (\omega_{s0})= \frac{\omega_{i0}}{\omega_{p0}}\left[n_{ie}(\omega_{i0})-n_{se}(\omega_{s0})\right]. \label{case4_1}
\end{eqnarray}
We find that for $\omega_{s0} \geq \omega_{i0}$, the right hand side of equation (\ref{case4_1}) is negative. Now in order for the left hand side of this equation to be negative, we have to ensure that $n_{po}(\omega_{p0}) < n_{ie} (\omega_{i0})$. Therefore, this scenario can be satisfied by positive uniaxial crystals.
\linebreak
\\
{\bf Case V: $\bm{ e \rightarrow o + e :}$}\\
Equation (\ref{n_eq}) in this case can be written as 
\begin{eqnarray}
n_{pe}(\omega_{p0})  - n_{so} (\omega_{s0})=  \frac{\omega_{i0}}{\omega_{p0}}\left[n_{ie}(\omega_{i0})-n_{so}(\omega_{s0})\right].
\end{eqnarray}
We take the SPDC to be degenerate, that is,  $\omega_{p0} = 2\omega_{s0}= 2\omega_{i0} $ and assume that the dispersion relation is linear for both the extraordinary and ordinary polarizations. We thus have $n_{pe}(2 \omega_{s0}) \approx 2 n_{pe}(\omega_{s0})$. Now, substituting $n_{pe}(\omega_{s0})=n_{ie}(\omega_{s0})=n_e(\omega_{s0})$ and $n_{so}(\omega_{s0})=n_o(\omega_{s0})$, we write the above equation as 
\begin{equation}
3 n_e(\omega_{s0})\approx n_o(\omega_{s0}).
\end{equation}
This equation holds only when $n_e < n_o$, that is, when the crystal is negative uniaxial.
\linebreak
\\
{\bf Case VI: $\bm {e \rightarrow e + o :}$}\\
Equation (\ref{n_eq}) in this case can be written as 
\begin{equation}
n_{pe}(\omega_p) - n_{se}(\omega_{s0}) =\frac{\omega_{i0}}{\omega_{p0}}\left[ n_{io}(\omega_{i0}) - n_{se}(\omega_{s0}) \right].
\end{equation}
Considering the same assumptions as in the above case, we write the above equation as 
\begin{equation}
3 n_e(\omega_{s0})- n_o(\omega_{i0}) \simeq 0,
\end{equation}
This equation holds only when $n_e < n_o$, that is, when the crystal is negative uniaxial.
\linebreak
\\
{\bf Case VII: $\bm{ o \rightarrow e + o :}$}\\
Equation (\ref{n_eq}) in this case can be written as 
\begin{equation}
n_{po}(\omega_{p0}) - n_{se}(\omega_{s0}) = \frac{\omega_{i0}}{\omega_{p0}}\left[ n_{io}(\omega_{i0}) - n_{se}(\omega_{s0}) \right].
\end{equation}
Considering the same assumptions as in the above case, we write the above equation as 
\begin{equation}
3 n_o(\omega_{s0})- n_e(\omega_{s0}) \simeq 0.
\end{equation}
This equation holds only when $n_o < n_e$, that is, when the crystal is positive uniaxial.
\linebreak
\\
{\bf Case VIII: $\bm {o \rightarrow o + e:}$}\\
Equation (\ref{n_eq}) in this case can be written as 
\begin{equation}
n_{po}(\omega_{p0}) - n_{so}(\omega_{s0}) = \frac{\omega_{i0}}{\omega_{p0}}\left[ n_{ie}(\omega_{i0}) - n_{so}(\omega_{s0}) \right].
\end{equation}
Considering the same assumptions as in the above case, we write the above equation as 
\begin{equation}
3 n_o(\omega_{s0})- n_e(\omega_{s0}) \simeq 0.
\end{equation}
This equation holds only when $n_o < n_e$, that is, when the crystal is positive uniaxial. 

In general, when both the signal and idler photons have the same polarization, it is referred to as the type-I phase matching and when they have orthogonal polarizations, it is referred to as the  type-II phase matching \cite{boeuf2000opteng,kurtsiefer2001jmo}. Thus, the possible polarization scenarios that can satisfy the phase-matching conditions can be represented in a tabular form as follows:
\begin{center}
\begin{tabular}{ |c|c|c| } 
 \hline
 Type & Positive uniaxial & Negative uniaxial \\ 
 \hline
 type-I &$ o \rightarrow e + e$ & $ e \rightarrow o + o$\\ 
 \hline
 type-II & $ o \rightarrow o + e $ & $ e \rightarrow e + o $ \\ 
  type-II & $ o \rightarrow e + o $ & $ e \rightarrow o + e $ \\ 
 \hline
\end{tabular}
\end{center}
Thus for positive uniaxial crystals, the pump field polarization has to be ordinary in order to satisfy the phase-matching condition. Similarly, for negative uniaxial crystals, the pump field polarization needs to be extraordinary in order to satisfy the phase matching condition.

\subsection{Wave propagation inside the SPDC crystals}

\begin{figure}[t!]
\centering
\includegraphics[scale=0.8]{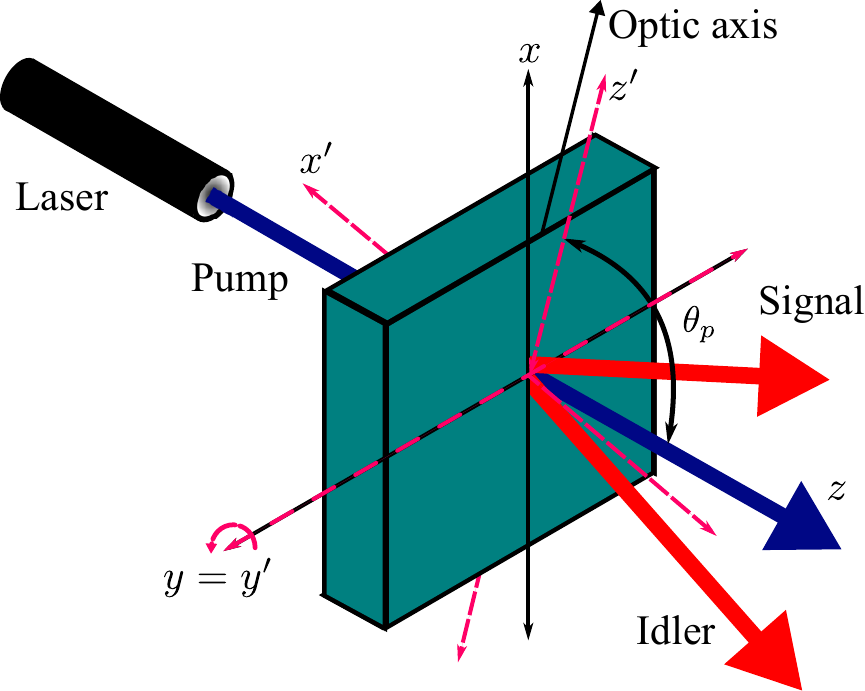}
\caption{The coordinate representation of the nonlinear crystal in both lab frame $(x,y,z)$ and rotated frame $(x', y',z')$. The rotated coordinate system is rotated by an angle $\theta_p$ around the y-axis.}
\label{fig:angle}
\end{figure}

In the previous subsection, we found out all possible polarization scenarios that can satisfy the phase-matching condition. In this subsection, we study how polarization affects the wave propagation inside an SPDC crystal. We note that in what follows we restrict our analysis only to negative uniaxial crystals and for brevity we represent the refractive indices without their frequency arguments. We start by writing the electric displacement vector $\bm D$ in terms of the electric field $\bm E$ inside an electrically anisotropic medium \cite{ghatak&thyagarajan1989}:
\begin{equation}\label{eq:29}
\left[{\begin{array}{c}
 D_x \\ 
 D_y\\
D_z
\end{array}}\right]
   = 
\left[{\begin{array}{ccc}
 \epsilon_{xx} &  \epsilon_{xy} &  \epsilon_{xz} \\ 
 \epsilon_{yx} &  \epsilon_{yy} &  \epsilon_{yz}\\
\epsilon_{zx} &  \epsilon_{zy} &  \epsilon_{zz}
 \end{array}}\right]
\left[{\begin{array}{c}
 E_x \\ 
 E_y\\
E_z
 \end{array}}\right].
\end{equation} 
Here, $(x,y,z)$ is the Cartesian coordinate system and $D_x$ and $E_x$ are the $x$-component of the electric displacement vector and electric field, etc. The 3$\times$3 matrix is known as the dielectric tensor with $\epsilon_{ij}$ being its elements, where $i, j=x, y, z$. Suppose the dielectric tensor can be written in the diagonal form in the $(x',y',z')$  coordinate system, that is, 
\begin{equation}\label{eq:30}
\hspace{20mm}\overline{\overline{ \epsilon} } = 
\left[{\begin{array}{ccc}
 \epsilon_{x'} &  0 &  0 \\ 
   0 &  \epsilon_{y'} &  0\\
 0 &  0  &  \epsilon_{z'}
 \end{array}}\right],
\end{equation}
where $(x', y', z')$ denotes the new coordinate system, which is obtained by rotating the $(x,y,z)$  coordinate system  by an angle $\theta_p$ around the $y$-axis as shown in figure (\ref{fig:angle}). The two coordinate systems are related as
\begin{equation}\label{eq:44}
\left[{\begin{array}{c}
 x' \\
 y'\\
z'
 \end{array}}\right] =
\left[{\begin{array}{ccc}
\cos \theta_p & 0 & \sin\theta_p \\ 
 0 & 1 &0\\
-\sin\theta_p & 0 & \cos\theta_p
 \end{array}}\right]
\left[{\begin{array}{ccc}
 x \\ 
 y\\
 z
\end{array}}\right].
\end{equation}
Next, we write $\epsilon_{z'}/\epsilon_0 = n^2_e$ and  $\epsilon_{x'}/\epsilon_0 = \epsilon_{y'}/\epsilon_0=n^2_o$ and express the electric displacement vector of equation (\ref{eq:29}) in the new coordinate system as
\begin{equation}\label{eq:31}
\left[{\begin{array}{c}
 D_{x'} \\ 
 D_{y'}\\
D_{z'}
 \end{array}}\right]
   = \epsilon_0
\left[{\begin{array}{ccc}
 n^2_0 &  0 &  0 \\ 
 0 &  n^2_0 &  0 \\
0  & 0  &  n^2_e
 \end{array}}\right]  
\left[{\begin{array}{c}
 E_{x'} \\ 
 E_{y'}\\
E_{z'}
\end{array}}\right].
\end{equation}
We know that a monochromatic electric field ${\bm E(\bm r,t)}$  propagating through an anisotropic medium can be written as
\begin{equation}\label{eq:32}
\qquad {\bm E({\bm r},t)} = {\bm E_0(\bm r)}\hspace{1mm} \mbox{exp}\left\lbrace i({\bm k}.{\bm r} - \omega t)\right\rbrace ,
\end{equation} 
where ${\bm k} = q_x \hat{ x } + q_y \hat{ y } + k_z \hat{ z } $ and ${\bm r} = x\hat{ x } + y\hat{ y }+ z\hat{ z } $. In the $(x', y',z')$ coordinate system, the field is given by
\begin{equation}\label{eq:34}
\qquad {\bm E}(\bm r',t) = {\bm E}_0(\bm r') \hspace{1mm} \mbox{exp}\left\lbrace i({\bm k}'.{\bm r}' - \omega t)\right\rbrace, 
\end{equation}
where  ${\bm k}' = q_{x'} \hat{ x' } + q_{y'} \hat{ y' } + k_{z'} \hat{ z' } $ and ${\bm r}' = x'\hat{ x' } + y'\hat{ y' }+ z'\hat{ z' } $. The wave equation in the $(x', y',z')$ coordinate system can be written as
\begin{equation}\label{eq:35}
\qquad {\bm\nabla'}({\bm \nabla'}\cdot{\bm E}) - {\bm \nabla'}^2 {\bm E} = -\mu_0 \frac{\partial^2 {\bm D}}{\partial t^2},  
\end{equation}
where ${\bm\nabla'}= \hat{ x' }\frac{\partial}{\partial x'} + \hat{ y' }\frac{\partial}{\partial y'}+ \hat{ z' }\frac{\partial}{\partial z'}$. Using equation (\ref{eq:31}), we write the above wave equation as
\begin{equation}\label{eq:36}
\qquad ({\bm k'}\cdot{\bm E}){\bm k}' - |{\bm k}'|^2 {\bm E} = -\left(\frac{\omega}{c}\right)^2 \left(\frac{\overline{ \overline{ \epsilon } }}{\epsilon_0}\right){\bm E}.
\end{equation} 
The above vector equation yields three simultaneous scalar equations involving the three components of the electric field:
\begin{eqnarray*}
\left[\left(\frac{n_o \omega}{c}\right)^2 - q^2_{y'}- k^2_{z'}\right]E_{x'} + q_{x'}q_{y'}E_{y'} + q_{x'}k_{z'}E_{z'}=0, \label{eq:37} \\
q_{x'} q_{y'}E_{x'} +\left[\left(\frac{n_o \omega}{c}\right)^2 - q^2_{x'}- k^2_{z'}\right]E_{y'} + q_{y'}k_{z'}E_{z'}=0 ,\label{eq:38}\\
q_{x'}k_{z'}E_{x'}+ q_{y'}k_{z'}E_{y'} +\left[\left(\frac{n_e \omega}{c}\right)^2 - q^2_{x'}- q^2_{y'}\right]E_{z'} =0. \label{eq:39}
\end{eqnarray*}
A non-trivial solution for $E_{x'}$, $E_{y'}$, $E_{z'}$ exists only if the determinant of the coefficients becomes zero, that is, only if
\begin{eqnarray}\label{eq:41}
&\left(\frac{\omega}{c}\right)^2\left[ \left( \frac{\omega}{c} n_o \right)^2 - q_{x'}^2 - q_{y'}^2 - k^2_{z'}\right]\times  \nonumber \\
&\left[ \left(\frac{\omega}{c} n_o n_e \right)^2 - (n_oq_{x'})^2 -(n_o q_y')^2 - (n_ek_{z'})^2\right] =0.
\end{eqnarray}
Since $\omega/c \neq 0$, the above condition requires that either
\begin{equation}\label{eq:42}
\frac{q^2_{x'} + q^2_{y'} + k^2_{z'} }{n^2_o} = \frac{\omega^2}{c^2},
\end{equation} 
or
\begin{equation}\label{eq:43}
 (n_oq_{x'})^2 +(n_o q_{y'})^2 + (n_ek_{z'})^2 - (\frac{\omega}{c} n_o n_e)^2=0.
\end{equation}
We note that the relation given in equation (\ref{eq:42}) is the dispersion relation for ordinary polarized light field and the relation given in equation (\ref{eq:43}) is the dispersion relation for the extraordinary polarized light field. Now, we need to write both these dispersion relations in the $(x, y, z)$ coordinate system. We note that the phase of a plane wave remains invariant under coordinate transformation, that is, ${\bm k}\cdot{\bm r}={\bm k}'\cdot{\bm r}'$, which in the component form can be written as
\begin{eqnarray}\label{eq:45}
(q_x x + q_y y + k_z z) = q_{x'}(x \cos\theta_p + z \sin\theta_p) + q_{y'}y \nonumber\\
\hspace{20mm} + k_{z'}(-x \sin\theta_p + z \cos\theta_p).
\end{eqnarray}
Comparing the coefficients of $x$, $y$, $z$ on both sides of the above equation, we obtain
\begin{eqnarray}
\qquad q_x = q_{x'}\cos\theta_p - k_{z'}\sin\theta_p, \label{eq:46}\\
\qquad q_y = q_{y'},\label{eq:47}\\
\qquad k_z = q_{x'}\sin\theta_p + k_{z'}\cos\theta_p.  \label{eq:48}
\end{eqnarray}
These relations can be inverted to yield
\begin{eqnarray}
\qquad q_{x'} = q_{x}\cos\theta_p + k_{z}\sin\theta_p,\label{eq:49}\\
\qquad q_{y'} = q_{y},\label{eq:50}\\
\qquad k_{z'} = -q_{x}\sin\theta_p + k_{z}\cos\theta_p. \label{eq:51}
\end{eqnarray}
Using the above relations, we write equation (\ref{eq:42}) in the $(x, y, z)$ coordinate system as
\begin{equation} \label{eq:52}
\qquad \frac{q^2_{x} + q^2_{y} + k^2_{z} }{n^2_o} = \frac{\omega^2}{c^2}.
\end{equation}
Therefore,  within the paraxial approximation, the longitudinal wave vector component $k_z$ of an ordinary polarized light field inside an anisotropic medium can be written as \cite{walborn2010pr}
\begin{equation} \label{eq:53}
k_z = \sqrt[]{\left(n_o \frac{\omega}{c}\right)^2 - q_x^2-q_y^2 } \approx n_o \frac{\omega}{c} - \frac{c}{2 n_o \omega}(q_x^2+q_y^2).
\end{equation} 
Similarly, using equations (\ref{eq:49}) through (\ref{eq:51}), we write equation (\ref{eq:43}) in the $(x, y, z)$ coordinate system as
\begin{eqnarray} \label{eq:54}
 \left[  \frac{\sin^2 \theta_p}{n^2_e } +  \frac{ \cos^2 \theta_p}{n^2_o }\right]  k^2_z  +   \left[  \frac{\cos^2 \theta_p}{n^2_e } +  \frac{ \sin^2 \theta_p}{n^2_o }\right]  q^2_x+ \frac{q^2_y}{n^2_e} \nonumber\\
\hspace{10mm} +\left[ \frac{1}{n^2_e} - \frac{1}{n^2_o}\right] \sin2\theta_p q_x k_z - \left(\frac{\omega}{c}\right)^2 =0.
\end{eqnarray}
The above equation is in the form of $a k^2_z + bk_z+c =0$. Therefore, solving for $k_z$ and using the paraxial approximation, the longitudinal wave vector component $k_z$ of the extraordinary polarized light field inside an anisotropic medium can be written as \cite{walborn2010pr}
\begin{eqnarray}
k_z &= -\alpha q_x + \sqrt[]{\left(\eta \frac{\omega}{c}\right)^2 -\beta^2 q^2_x - \gamma^2 q^2_y}, \nonumber \\
&\approx -\alpha q_x + \eta \frac{\omega}{c} - \frac{c}{2 \eta \omega}\left[ \beta^2 q^2_x + \gamma^2 q^2_y\right], \label{eq:55}
\end{eqnarray}
where 
\begin{eqnarray}
\hspace{20mm}\alpha = \frac{(n^2_o- n^2_e)\sin\theta_p \cos\theta_p}{n^2_o\sin^2\theta_p + n^2_e \cos^2 \theta_p}, \label{eq:56}\\
\hspace{20mm}\beta = \frac{n_o n_e}{n^2_o\sin^2\theta_p + n^2_e \cos^2 \theta_p},\label{eq:57}\\
\hspace{20mm}\gamma = \frac{n_o }{\sqrt[]{n^2_o\sin^2\theta_p + n^2_e \cos^2 \theta_p}},\label{eq:58}\\
\hspace{20mm}\eta = \frac{n_o n_e }{\sqrt[]{n^2_o\sin^2\theta_p + n^2_e \cos^2 \theta_p}}. \label{eq:59}
\end{eqnarray}
Here $\theta_p$ is the angle between the optic axis and the pump propagation direction, that is, the $z$-axis. By changing the angle $\theta_p$, one can generate entangled photons with various different phase matching conditions.

\section{Calculating the detection probabilities of the generated entangled photons}\label{det-prob}

In this section, we derive explicit expressions for the phase-matching function for various possible polarization scenarios. We restrict our analysis to the BBO crystal which is a negative uniaxial crystal. We note that the dispersion relations for the ordinary and extraordinary polarized light inside the BBO crystal are given by \cite{eimerl1987jap} 
\begin{eqnarray}
n^2_o=  2.7405 + \frac{0.0184}{\lambda^2 -0.0179} - 0.0155 \lambda^2,\\
n^2_e= 2.3730 + \frac{0.0128}{\lambda^2 -0.0156} -0.0044\lambda^2.
\end{eqnarray}
Here $\lambda$ is the  wavelength of light in $\mu$m. Also, for brevity, we show the refractive indices without their corresponding frequency arguments.

\subsection{Calculating the coincidence and the individual photon count rates}

In this article, we are primarily interested in studying the two-photon wavefunction in the position and linear momentum bases. Therefore, in what follows, we assume that the signal, idler and pump fields are monochromatic with central frequencies given by $\omega_{s0}$, $\omega_{i0}$, and $\omega_{p0}$, respectively. With this assumption and using equation (\ref{psitp_detctor}), we write the state of the down-converted two-photon field at the exit face of the crystal as
\begin{figure}[t!]
\centering
\includegraphics[scale=0.8]{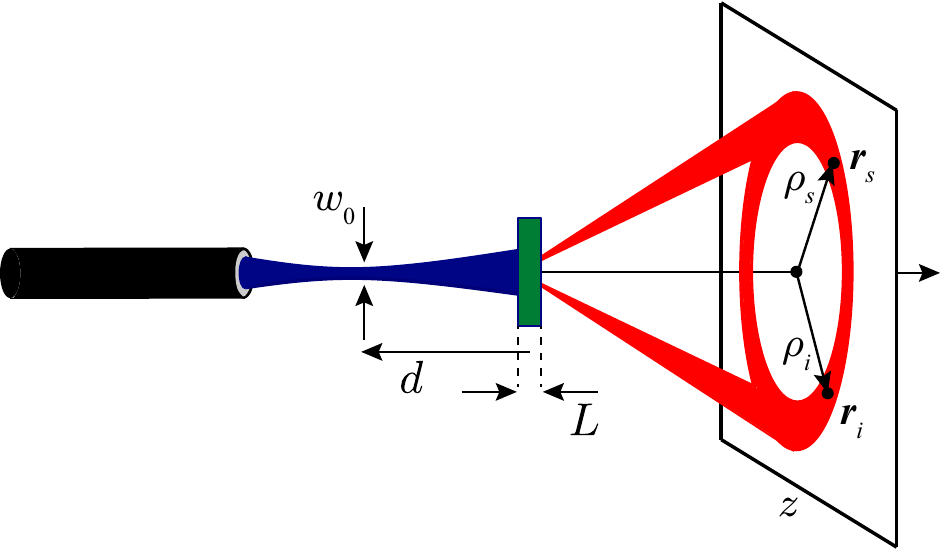}
\caption{Experimental set up for the generation of SPDC photon pairs. The pump beam waist $w_o$ is located at a distance $d$ from the front face of the crystal. The photons are detected at a distance $z$ from the crystal. }
\label{field_propagation}
\end{figure}
\begin{eqnarray}\label{psitp}
|\psi_{\rm tp}\rangle =A \int \!\!\!\!\int d^2 {\bm q}_s d^2 {\bm q}_i V({\bm q}_s+{\bm q}_i) \Phi({\bm q}_s,{\bm q}_i)  \ket{{\bm q}_s}_s \ket{{\bm q}_i}_i. \nonumber\\
\end{eqnarray}
Here, we do not explicitly show the frequency arguments. 
We now calculate the probability of detecting the signal and idler photons at a distance $z$ from the crystal at locations $\bm r_s\equiv (\bm\rho_s, z)$ and $\bm r_i\equiv (\bm\rho_i, z)$, respectively, in coincidence (see figure \ref{field_propagation}). The coincidence count rate $R_{si}(\bm r_{s}, \bm r_{i})$, which is the probability per (unit time)$^{2}$ that a signal photon is detected at position $(\bm\rho_s, z)$ and an idler photon is detected at position
$(\bm\rho_i, z)$, is given by \cite{glauber1963pra}:
\begin{eqnarray}
R_{si}(\bm r_{s}, \bm
r_{i})= R_{si}(\bm\rho_s, \bm\rho_i, z),\nonumber \\ \quad =\alpha_{s}\alpha_{i} \langle \psi_{\rm tp}|
\hat{E}_{s}^{(-)}(\bm{r}_{s})\hat{E}_{i}^{(-)}(\bm{r}_{i})
\hat{E}_{i}^{(+)}(\bm{r}_{i})
\hat{E}_{s}^{(+)}(\bm{r}_{s})|\psi_{\rm
tp} \rangle, \nonumber \\
\quad  = \alpha_{s}\alpha_{i} \Big| \langle {\rm vac}|_s\langle {\rm vac}|_i|\hat{E}_{s}^{(+)}(\bm{r}_{s})\hat{E}_{i}^{(+)}(\bm{r}_{i})|\psi_{\rm
tp} \rangle \Big|^2,  \label{coin_eqn}
\end{eqnarray}
Here $\alpha_{s}$ and $\alpha_{i}$ denote the quantum
efficiencies of the detectors kept at locations $(\bm\rho_s, z)$ and $(\bm\rho_i, z)$, respectively. $\hat{E}^{(+)}_{s}({\bm r}_{s})$ and $\hat{E}^{(+)}_{i}({\bm r}_{i})$ denote the positive-frequency parts of the electric field operators at detection locations $ (\bm\rho_s, z)$ and $ (\bm\rho_i, z)$, respectively. The two field operators can be written as
\begin{eqnarray}
\hat{E}_{s}^{(+)}(\bm{r}_{s})=\hat{E}_{s}^{(+)}(\bm{\rho}_{s}, z), \nonumber \\
\qquad =\int d^2\bm{q}_s  e^{ik_{s}z}
e^{i(\bm{q}_s\cdot\bm{\rho}_{s}-q_s^{2}z/2k_{s})}\hat{a}(\bm{q}_s),  \label{field A}
\end{eqnarray}
\begin{eqnarray}
\hat{E}_{i}^{(+)}(\bm{r}_{i})=\hat{E}_{i}^{(+)}(\bm{\rho}_{i}, z), \nonumber \\
\qquad = \int d^2\bm{q}_i e^{ik_{i}z}
e^{i(\bm{q}_i\cdot\bm{\rho}_{i}-q_i^{2}z/2k_{i})} \hat{a}(\bm{q}_i).  \label{field B} 
\end{eqnarray}
Here $q_s^{2}=|\bm{q}_s|^{2}$, $q_i^{2}=|\bm{q}_i|^{2}$,
$k_s=|\bm k_s|= \omega_{s0}/c$, and $k_i=|\bm k_i|= \omega_{i0}/c$, where $c$ is the speed of light. We note that in writing the electric field operators $\hat{E}_{s}^{(+)}(\bm{\rho}_{s},
z)$ and $\hat{E}_{i}^{(+)}(\bm{\rho}_{i}, z)$ above, we have made use of the paraxial approximation. Substituting from equations (\ref{psitp}), (\ref{field A}) and (\ref{field B}) in equation (\ref{coin_eqn}), we obtain the following expression for the coincidence count rate:
\begin{eqnarray}
R_{si}(\bm\rho_s, \bm\rho_i, z)= \alpha_s \alpha_i \Bigg|A  \int \!\!\!\!\int \!\!\!\!\int \!\!\!\!\int d^2 {\bm q}_s d^2 {\bm q}_i d^2 {\bm q}'_s d^2 {\bm q}'_i \nonumber \\
\qquad\quad\times e^{i(k_s + k_i)z} V({\bm q}_s + {\bm q}_i)\Phi({\bm q}_s, {\bm q}_i) \nonumber\\
\qquad\quad\times {\rm exp}\left[ {i\left({\bm q}'_s\cdot {\bm \rho}_s + {\bm q}'_i\cdot {\bm \rho}_i - \frac{|{\bm q}_s'|^2 z}{2 k_s} -  \frac{|{\bm q}_i'|^2 z}{2 k_i}\right)}\right]  \nonumber \\
\qquad\quad \times \langle{\rm vac}|_s\langle{\rm vac}|_i\hat{a}(\bm{q}'_s) \hat{a}(\bm{q}'_i) |{\bm q}_s\rangle_s|{\bm q}_i\rangle_i \Bigg|^2. \label{coin_eqn2}
\end{eqnarray}
Now using the relation $\hat{a}(\bm{q}'_s) \hat{a}(\bm{q}'_i) |{\bm q}_s \rangle_s|{\bm q}_i,\rangle_i = \delta({\bm q}'_s - {\bm q}_s)\delta({\bm q}'_i - {\bm q}_i)|{\rm vac}\rangle_s |{\rm vac}\rangle_i$,  we write equation (\ref{coin_eqn2}) as 
\begin{eqnarray}\label{coincidence}
R_{si}(\bm\rho_s, \bm\rho_i, z)= \alpha_s \alpha_i \Bigg|A  \int \!\!\!\!\int d^2 {\bm q}_s d^2 {\bm q}_i~ e^{i(k_s + k_i)z} \nonumber \\
 \qquad\times V({\bm q}_s + {\bm q}_i)\Phi({\bm q}_s, {\bm q}_i) \nonumber\\
\quad\times {\rm exp}\left[ {i\left({\bm q}_s\cdot {\bm \rho}_s + {\bm q}_i\cdot {\bm \rho}_i - \frac{|{\bm q}_s|^2 z}{2 k_s} -  \frac{|{\bm q}_i|^2 z}{2 k_i}\right)}\right] \Bigg|^2 . 
\end{eqnarray}
In addition to the coincidence count rate, we also calculate the photon count rates of the individual signal and idler photons at locations $ (\bm\rho_s, z)$ and $ (\bm\rho_i, z)$, respectively. The photon count rate at the signal location $ (\bm\rho_s, z)$ is obtained by integrating the coincidence count rates over all possible idler locations, and vice versa. Thus the signal and idler photon count rates at locations $(\bm\rho_s, z)$ and $ (\bm\rho_i, z)$, respectively, can be written as 
\begin{eqnarray}
R_s(\bm\rho_s, z)= \int R_{si}({\bm \rho}_s, {\bm \rho}_i,z)d^2\bm\rho_i , \label{signal count} \\
R_i(\bm\rho_i, z)= \int R_{si}({\bm \rho}_s, {\bm \rho}_i,z)d^2\bm\rho_s . \label{idler count}
\end{eqnarray}

\subsection{Calculating the conditional detection probability}

In addition to the coincidence and individual photon count rates, in many experimental situations, one is also interested in the conditional detection probabilities of individual photons. For example, it may of interest in certain experiment to calculate the detection probability of the signal photon as a function of position ${\bm \rho}_{s}$ given that the idler photon is already detected at position ${\bm \rho}_{i0}$. In fact, quite often, by detecting an idler at a particular position the region over which the signal photon could be found becomes very localized. The width of this region is called the position-correlation width. The position-correlation of entangled photons have been studied and utilized in several experimental situations \cite{grayson1994pra,joobeur1994pra,strekalov1995prl,ribeiro1994pra}. The position-correlation width is calculated as follows. For the fixed idler position $({\bm \rho_{i0}},z)$ the coincidence count rate can be written using equation (\ref{coincidence}) as

\begin{eqnarray}\label{conditional_probability_eqn}
R_{si}({\bm \rho}_s, {\bm \rho}_{i0},z)= \alpha_s \alpha_i \Bigg|A  \int \!\!\!\!\int d^2 {\bm q}_s d^2 {\bm q}_i~ e^{i(k_s + k_i)z} \nonumber \\
\times  V({\bm q}_s + {\bm q}_i)\Phi({\bm q}_s, {\bm q}_i) \nonumber\\
\times {\rm exp}\left[ {i\left({\bm q}_s\cdot {\bm \rho}_s + {\bm q}_i\cdot {\bm \rho}_{i0} - \frac{|{\bm q}_s|^2 z}{2 k_s} -  \frac{|{\bm q}_i|^2 z}{2 k_i}\right)}\right] \Bigg|^2 ,
\end{eqnarray}
where ${\bm r}_s = ({\bm \rho}_s , z)$ and  ${\bm r}_{i0} = ({\bm \rho}_{i0} , z)$. The width of  $R_{si}({\bm \rho}_s, {\bm \rho}_{i0},z)$  is the position-correlation width of the signal photon at a plane $z$ away from the crystal. We note that in a similar manner, one can also calculate the momentum-correlation width.

Equation (\ref{coincidence}) for the coincidence count rate and equations (\ref{signal count}) and (\ref{idler count}) for photon count rates at the signal and idler locations have been derived for a very generic pump field and the phase matching function. We now calculate the phase-matching function $\Phi(\bm q_s, \bm q_i)$ for both type-I and type-II phase-matching conditions. Using the  expression for $\Phi(\bm q_s, \bm q_i)$ in a particular phase-matching condition, one can calculate the coincidence count rate using equation~(\ref{coincidence}) and the individual photon count rates using equations (\ref{signal count}) and (\ref{idler count}).

\subsection{Calculating the detection probabilities for type-I Phase matching}
In type-I phase matching, the pump field is extraordinary polarised and the down-converted fields are ordinary polarised. Therefore, with paraxial approximation, the $z$-component of the pump, signal, and idler fields are given by 
\begin{eqnarray}
&k_{pz} = -\alpha_p q_{px} + \eta_p \frac{\omega_{p0}}{c} - \frac{c \left[ \beta_p^2 q^2_{px} + \gamma_p^2 q^2_{py}\right]}{2 \eta_p \omega_{p0}},  \label{eq:60}\\
&k_{sz} =  n_{so} \frac{\omega_{s0}}{c} - \frac{c}{2 n_{so} \omega_{s0}}(q_{sx}^2+q_{sy}^2), \label{eq:61}\\
&k_{iz} =  n_{io} \frac{\omega_{i0}}{c} - \frac{c}{2 n_{io} \omega_{i0}}(q_{ix}^2+q_{iy}^2).  \label{eq:62}
\end{eqnarray}
Here 
\begin{eqnarray*}
~~~~~~\alpha_p = \frac{(n^2_{po}- n^2_{pe})\sin\theta_p \cos\theta_p}{n^2_{po}\sin^2\theta_p + n^2_{pe} \cos^2 \theta_p}, \label{eq:56}\\
~~~~~~\beta_p = \frac{n_{po} n_{pe}}{n^2_{po}\sin^2\theta_p + n^2_{pe} \cos^2 \theta_p},\label{eq:57}\\
~~~~~~\gamma_p = \frac{n_{po} }{\sqrt[]{n^2_{po}\sin^2\theta_p + n^2_{pe} \cos^2 \theta_p}},\label{eq:58}\\
~~~~~~\eta_p = \frac{n_{po} n_{pe} }{\sqrt[]{n^2_{op}\sin^2\theta_p + n^2_{pe} \cos^2 \theta_p}}. \label{eq:59}
\end{eqnarray*}
The phase-matching function in this case is given by
\begin{eqnarray}\label{phase-matching-eoo}
\Phi({\bm q}_s,{\bm q}_i) &\rightarrow \Phi_{eoo}({\bm q}_s,{\bm q}_i) \nonumber \\ &=L~\mbox{sinc}\left[ \Delta k_z\frac{L}{2}\right] \mbox{exp} \left\lbrace i \Delta k_z \frac{L}{2}\right\rbrace  ,
\end{eqnarray} 
where 
\begin{eqnarray}\label{eq:delkz64}
&\Delta k_z = k_{sz} + k_{iz}- k_{pz},\nonumber\\
 &=  n_{os} \frac{\omega_{s0}}{c} +  n_{oi} \frac{\omega_{i0}}{c} - \eta_p \frac{\omega_{p0}}{c}  + \frac{c}{2 \eta_p \omega_{p0}}\left[ \beta_p^2 q^2_{px} + \gamma_p^2 q^2_{py}\right] \nonumber\\
& + \alpha_p(q_{sx}+q_{ix}) - \frac{c}{2 n_{os} \omega_{s0}}|{\bm q}_s|^2 -\frac{c}{2 n_{oi} \omega_{i0}}|{\bm q}_i|^2.
\end{eqnarray}
Here $\Phi_{eoo}({\bm q}_s,{\bm q}_i) $ represents the phase matching function when the pump field is extraordinary polarized while both the signal and idler fields are ordinary polarized.

\subsection{Calculating the detection probabilities for type-II Phase matching}
In type-II phase matching, the pump field is extraordinary polarized while the polarizations of the signal and idler fields are orthogonal to each other. Hence there are two possibilities for the signal and idler polarizations in this type of phase matching.\\
\linebreak
{\bf Case-1 :} If the signal is extraordinary polarized and the idler is ordinary, then we have
\begin{eqnarray}
&k_{pz} = -\alpha_p q_{px} + \eta_p \frac{\omega_{p0}}{c} - \frac{c \left[ \beta_p^2 q^2_{px} + \gamma_p^2 q^2_{py}\right]}{2 \eta_p \omega_{p0}}, \label{eq:65}\\
&k_{sz} =   -\alpha_s q_{sx} + \eta_s \frac{\omega_{s0}}{c} - \frac{c \left[ \beta_s^2 q^2_{sx} + \gamma_s^2 q^2_{sy}\right]}{2 \eta_s \omega_{s0}},  \label{eq:66}\\
&k_{iz} =  n_{io} \frac{\omega_{i0}}{c} - \frac{c}{2 n_{io} \omega_{i0}}|{\bm q}_i|^2. \label{eq:67}
\end{eqnarray}
The phase matching function in this case is given by
\begin{eqnarray} \label{phase-matching-eeo}
\Phi({\bm q}_s,{\bm q}_i) & \rightarrow \Phi_{eeo}({\bm q}_s,{\bm q}_i) \nonumber \\ &= L~\mbox{sinc}\left[ \Delta k_{1z}\frac{L}{2}\right] \mbox{exp} \left\lbrace i \Delta k_{1z} \frac{L}{2}\right\rbrace,
\end{eqnarray}
where
\begin{eqnarray}
\Delta k_{1z} &= k_{sz} + k_{iz}- k_{pz} \nonumber\\
 &=   -\alpha_s q_{sx} + \eta_s \frac{\omega_{s0}}{c} - \frac{c}{2 \eta_s \omega_{s0}}\left[ \beta_s^2 q^2_{sx} + \gamma_s^2 q^2_{sy}\right]   \nonumber \\ 
 &+  n_{io} \frac{\omega_{i0}}{c} - \frac{c}{2 n_{io} \omega_{i0}}|{\bm q}_i|^ 2  +\alpha_p(q_{sx}+q_{ix}) - \eta_p \frac{\omega_{p0}}{c} \nonumber \\
 & + \frac{c}{2 \eta_p \omega_{p0}}\left[ \beta_p^2(q_{sx}+q_{ix})^2 + \gamma_p^2(q_{sy}+q_{iy})^2\right] . \label{eq:67}
\end{eqnarray}
Here $\Phi_{eeo}({\bm q}_s,{\bm q}_i) $ represents the phase matching function. We note that $\alpha_p,\beta_p, \gamma_p, \eta_p$ are calculated by putting in the values of $n_o$ and $n_e$ at the pump frequency $\omega_{p0}$ while $\alpha_s,\beta_s, \gamma_s, \eta_s$ are calculated by putting in the values of $n_o$ and $n_e$ at the signal frequency $\omega_{s0}$. The angle $\theta_p$ remains the same in each case.\\
\linebreak
{\bf Case-2 :}If the signal is extraordinary polarized and the idler is ordinary, then we have
\begin{eqnarray}
&k_{pz} = -\alpha_p q_{px} + \eta_p \frac{\omega_{p0}}{c} - \frac{c \left[ \beta_p^2 q^2_{px} + \gamma_p^2 q^2_{py}\right]}{2 \eta_p \omega_{p0}}, \\
&k_{sz} =  n_{so} \frac{\omega_{s0}}{c} - \frac{c}{2 n_{so} \omega_{s0}}|{\bm q}_s|^2,\\
&k_{iz} =   -\alpha_i q_{ix} + \eta_i \frac{\omega_{i0}}{c} - \frac{c \left[ \beta_i^2 q^2_{ix} + \gamma_i^2 q^2_{iy}\right]}{2 \eta_i \omega_{i0}}. 
\end{eqnarray}
The phase matching function in this case is given by
\begin{eqnarray} \label{phase-matching-eoe}
\Phi({\bm q}_s,{\bm q}_i) & \rightarrow \Phi_{eoe}({\bm q}_s,{\bm q}_i) \nonumber \\ & = L~\mbox{sinc}\left[ \Delta k_{2z}\frac{L}{2}\right] \mbox{exp} \left\lbrace i \Delta k_{2z} \frac{L}{2}\right\rbrace,
\end{eqnarray}
where
\begin{eqnarray}
\Delta k_{2z} &= k_{sz} + k_{iz}- k_{pz}, \nonumber\\
 &= n_{so} \frac{\omega_{s0}}{c} - \frac{c}{2 n_{so} \omega_{s0}}|{\bm q}_s|^2  -\alpha_i q_{ix} + \eta_i \frac{\omega_{i0}}{c} \nonumber \\
 & - \frac{c \left[ \beta_i^2 q^2_{ix} + \gamma_i^2 q^2_{iy}\right]}{2 \eta_i \omega_{i0}}  +\alpha_p (q_{sx}+q_{ix}) - \eta_p \frac{\omega_{p0}}{c}  \nonumber \\
 &+ \frac{c \left[ \beta_p^2(q_{sx}+q_{ix})^2 + \gamma_p^2(q_{sx}+q_{ix})^2\right]}{2 \eta_p \omega_{p0}}.  \label{eq:69}
\end{eqnarray}
Here $\Phi_{eoe}$ is the phase matching function for pump, signal and idler being  extraordinary, ordinary and extraordinary polarized light respectively. We note that $\alpha_p,\beta_p, \gamma_p, \eta_p$ are calculated by putting in the values of $n_o$ and $n_e$ at the pump frequency $\omega_{p0}$ while $\alpha_i,\beta_i, \gamma_i, \eta_i$ are calculated by putting in the values of $n_o$ and $n_e$ at the idler frequency $\omega_{i0}$. We note that in experimental situations, we get both the above cases of type-II  phase matching satisfied at the same time. Therefore, the coincidence and the individual count rates have contributions due to both the phase-matching functions, $\Phi_{eoe}({\bm q}_s,{\bm q}_i)$ and $\Phi_{eeo}({\bm q}_s,{\bm q}_i)$.

\section{Experimental and numerical results}\label{result}

In this section, we present results of our experimental and numerical studies. We evaluate equations (\ref{coincidence}) through (\ref{idler count}) in order to study how the generation of entangled photons depend on various experimental parameters. In most cases, we also present our experimental observations in order to highlight how it matches with the theory. In our experiments, we use a $405$ nm, $100$ mW continuous wave laser as the pump field. The pump power
of 100 mW ensured that we were working within the weak down-conversion limit,
in which case the probability of producing a four-photon state is negligibly small
compared to that of producing a two-photon state.  The pump field is in the form of a Gaussian beam having a beam-waist of  $388\ {\rm \mu m}$. For the nonlinear optical crystal, we use a $2$ mm thick $\beta$-barium borate (BBO) crystal. By putting suitable wavelength filters, we make sure that we have degenerate down-conversion with the central wavelength of both the signal and idler photons being $810$ nm. The distance between the crystal and the beam waist location turns out to be $d= 107.8$ cm. For recording the down-converted photons, we use an Andor iXon Ultra
EMCCD camera having $512\times512$ pixels and kept at $z=35$ mm. For our numerical experiments, we use the same value of the experimental parameters as mentioned above.

\subsection{Effect of $\theta_p$ on the down-conversion output}

First of all we study how the angle $\theta_p$ between the optic axis and the pump propagation direction affects the down-conversion output. For a range of values of $\theta_p$, we numerically evaluate equation (\ref{coincidence}) and (\ref{signal count}) for various different values of $\theta_p$ and with both type-I and type-II phase-matching functions as given in equations (\ref{phase-matching-eoo}),  (\ref{phase-matching-eeo}), and (\ref{phase-matching-eoe}). We find that for $\theta_p < 28.6^{\circ}$, no phase-matching gets satisfied.

\begin{figure}[hbt!]
\centering
\includegraphics[scale=0.9]{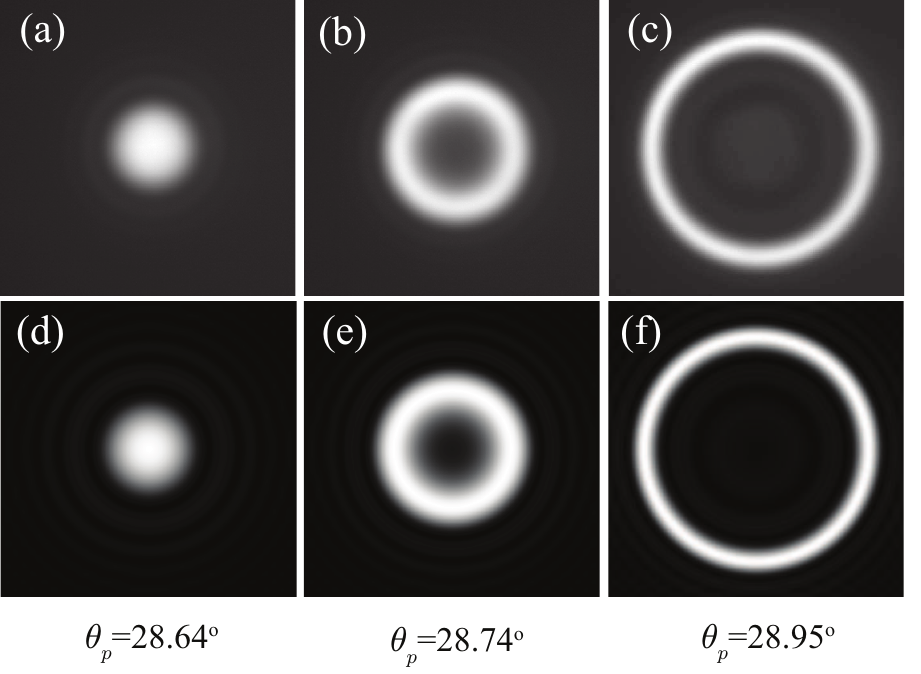}
\caption{(a), (b), (c) are the experimental signal photon intensity obtained using an EMCCD camera kept at $z=35$ mm for various different values of $\theta_p$. (a)  $\theta_p$= 28.64$^{\circ}, \alpha= 0.33^{\circ} $  (b) $\theta_p$= 28.74$^{\circ}, \alpha= 0.49^{\circ} $ (c) $\theta_p$= 28.95$^{\circ}, \alpha= 0.82^{\circ} $. The corresponding numerical plots are shown in (d), (e), and (f) respectively.}
\label{theta_variation_typeI}
\end{figure}

Figure \ref{theta_variation_typeI} shows the two-dimensional images of down-conversion output obtained using an EMCCD camera and also the numerically calculated two-dimensional plots $R(\bm\rho_s, z)$ for $z=35$ mm at various different values of $\theta_p$. Figures \ref{theta_variation_typeI}(a), \ref{theta_variation_typeI}(b) and \ref{theta_variation_typeI}(c) are the experimental plots  while \ref{theta_variation_typeI}(d), \ref{theta_variation_typeI}(e) and \ref{theta_variation_typeI}(f) are the corresponding numerical plots. We see that the collinear type-I phase matching starts around $\theta_p=28.6^{\circ}$ and the output pattern looks like a circular blob. When $\theta_p$ is increased further the circular blob turns into an annulus with non-collinear phase matching. The radius of the annulus increases as  $\theta_p$ is increased. We see a very good match between theory and experiment.

\begin{figure}[hbt!]
\centering
\includegraphics[scale=0.9]{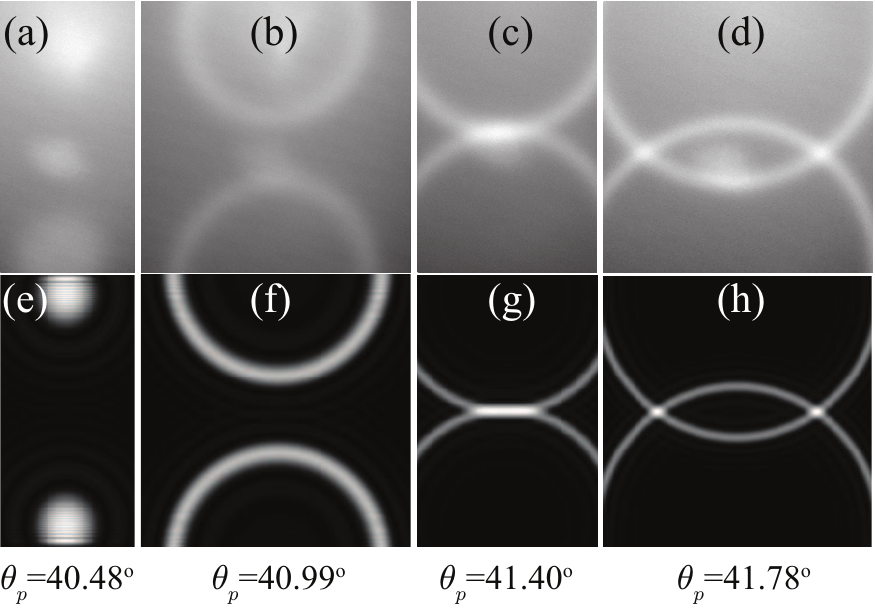}
\caption{(a), (b), (c), and (d) are the experimental signal photon intensity obtained using an EMCCD camera kept at $z=35$ mm for various different values of $\theta_p$. (a)  $\theta_p$= 40.48$^{\circ}, \alpha= 19.11^{\circ} $  (b) $\theta_p$= 40.99$^{\circ}, \alpha= 19.94^{\circ} $ (c) $\theta_p$= 41.40$^{\circ}, \alpha= 20.60^{\circ}$ (d) $\theta_p$= 41.78$^{\circ}, \alpha= 21.23^{\circ} $. The corresponding numerical plots are shown in (e), (f), (g), and (h) respectively. }
\label{SPDCII_theta}
\end{figure}
We find that the type-II phase matching starts around $\theta_p=40.48^{\circ}$ in the form of two off-axis blobs. Figure \ref{SPDCII_theta} shows the two-dimensional images of down-conversion output obtained using an EMCCD camera and also the numerically calculated two-dimensional plots $R(\bm\rho_s, z)$ for $z=35$ mm at various different values of $\theta_p$ for type-II phase matching. Figures \ref{SPDCII_theta}(a), \ref{SPDCII_theta}(b), \ref{SPDCII_theta}(c), and  \ref{SPDCII_theta}(d) are the experimental plots  while \ref{SPDCII_theta}(e), \ref{SPDCII_theta}(f), \ref{SPDCII_theta}(g), and  \ref{SPDCII_theta}(h) are the corresponding numerical plots. We note that at $\theta_p=40.48^{\circ}$ both the blobs have orthogonal polarizations and that one blob corresponds to the signal photon while the other one corresponds to the idler photon. As $\theta_p$ is increased, the two blobs open up and become two separate annuli. When $\theta_p$ is increased even further the two annuli start overlapping with each other. We find that the match between theory and experiment is not as good as in the case of type-I phase-matching. This is because of the fact that the BBO crystal that we are using in our experiments is meant to be used for type-I phase matching. Using it for type-II phase matching means rotating the crystal about the $y$-axis by large angles which causes a large background observed in our experiments. Nevertheless, as far as the overall $\theta_p$ dependence is concerned, the match between the theory and experiment seems quite good.

\subsection{Effect of $\theta_p$ on the total output power of the down-converted photons}

\begin{figure}[h!]
\centering
\includegraphics[scale=0.8]{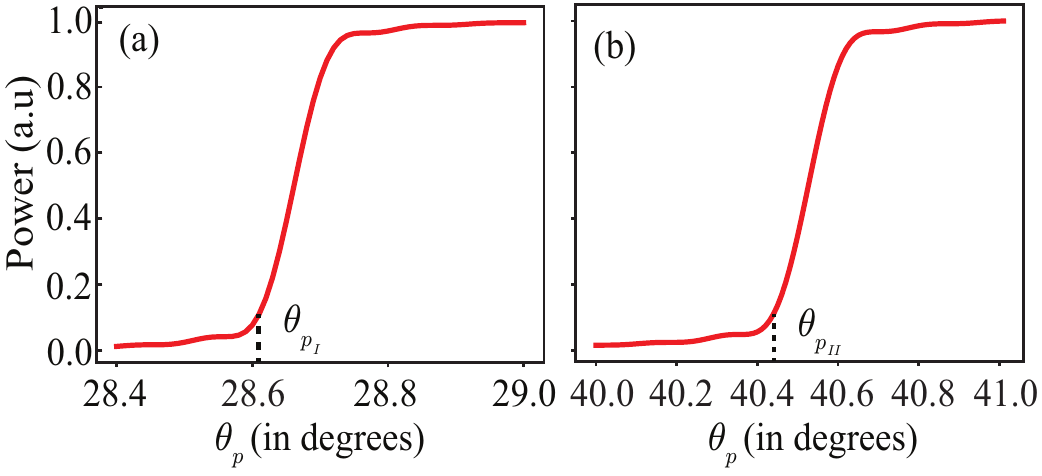}
\caption{Numerically calculated total output power as a function of $\theta_p$ for (a) type-I phase matching and (b) type-II phase matching. type-I phase matching takes place for  $\theta_p > 28.6^{\circ}$  while type-II phase matching takes place for $\theta_p > 40.48^{\circ}$.}
\label{intensity_vs_thetap}
\end{figure}

Next, we numerically study how the output power of the down-converted photons depend on the angle $\theta_p$ between the optic axis and the pump propagation direction. For this, we numerically evaluate equation (\ref{signal count}) and (\ref{idler count}) for various different values of $\theta_p$ such that it covers both type-I and type-II phase matching scenarios. For each value of $\theta_p$, we calculate the total output power of the down-converted photons by integrating over the two-dimensional space. In Figure \ref{intensity_vs_thetap}, we plot this power  as a function of $\theta_p$. While figure~\ref{intensity_vs_thetap}(a) shows this plot over $\theta_p$ values that give type-I phase matching,   figure~\ref{intensity_vs_thetap}(b) shows this plot over  $\theta_p$ values that give type-II phase matching. 

We note that in  figure \ref{intensity_vs_thetap}(a) the output power is nearly zero for $\theta_p < 28.6^{\circ}$. This means that for  $\theta_p< 28.6^{\circ}$, the phase matching condition is not satisfied. We get type-I phase matching for $\theta_p > 28.6^{\circ}$. We note that once the total output power reaches its maximum, a further increase in $\theta_p$ value does not change the output power, which then becomes almost constant as a function of $\theta_p$. In figure \ref{intensity_vs_thetap}(b), the total output power is nearly zero for $\theta_p < 40.48 ^{\circ}$. Here we note that for $\theta_p > 28.6 ^{\circ}$, type-I phase matching is always satisfied. However, the size of the anuulus becomes very large. So, for $\theta_p > 40.48 ^{\circ}$, we calculate the contribution to the total power only due to the type-II phase matching. We find that as for $\theta_p > 40.48 ^{\circ}$, the total output power due to type-II phase matching starts to increase and once it reaches its maximum value it stays constant as a function for $\theta_{p}$. Thus for both type-I and type-II phase matching, we find that the total output power after reaching its maximum value becomes independent of $\theta_p$, even though the output intensity pattern changes.

\subsection{Variation of type-I intensity with $\theta_p$}

\begin{figure}[t!]
\centering
\includegraphics[scale=0.8]{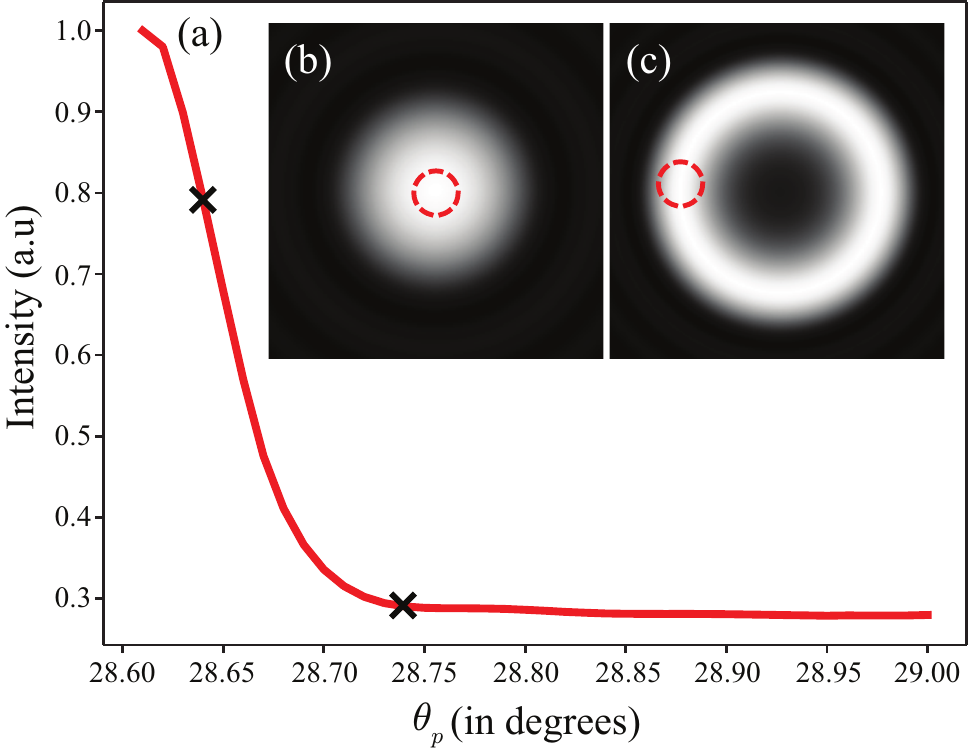}
\caption{(a) Numerically calculated maximum intensity of signal photons as a function of $\theta_p$ for type-I phase matching condition at $z=$ 35 mm. Collinear to non-collinear phase matching is achieved by increasing $\theta_p$ angle. (b) The intensity profile obtained with collinear phase matching at $\theta_p= 28.64^{\circ}$ (c) The intensity profile obtained at $\theta_p= 28.74^{\circ}$. Dashed circles are the area over which the intensity is calculated. }
\label{col_noncol_intensity}
\end{figure}

In this subsection, we study how the intensity of down-converted photons changes as a function of $\theta_p$ for the type-I phase matching condition. For this purpose, first we numerically calculate the signal photon intensity $R(\bm\rho_s, z)$ for various different values of $\theta_p$. Then, as shown in figure \ref{col_noncol_intensity}, we select small regions on the intensity plots to find the maximum intensity. Figure \ref{col_noncol_intensity} (a) shows the variation of intensity as a function of $\theta_p$. We find that the photons have maximum intensity in the collinear emission geometry. As the phase matching becomes more and more non-collinear, the intensity becomes lesser. This result shows that the intensity and thus the photon-emission efficiency is higher in collinear phase matching compared to the non-collinear phase matching. Thus, in experiments in which more intense photon source is required the collinear phase matching should be preferred. Here, as illustrated in the previous section, we note that the total output power remains the same irrespective of whether the collinear or the non-collinear phase matching is satisfied. 

\subsection{Effect of crystal length on down-conversion}

\begin{figure}[t!]
\centering
\includegraphics[scale=0.9]{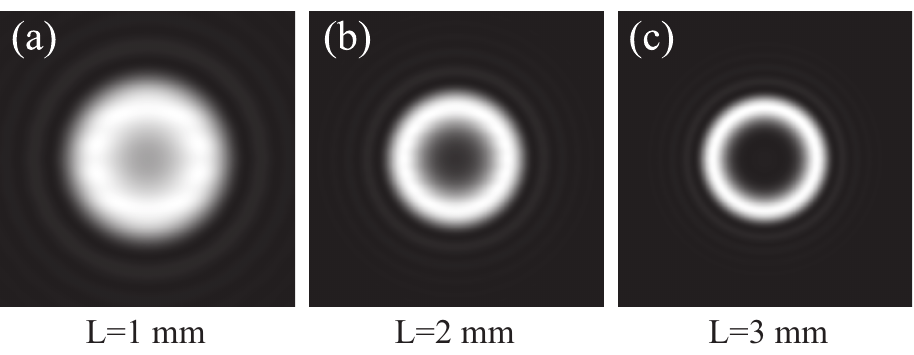}
\caption{Numerically simulated down-conversion intensity patterns for type-I phase matching at different crystals lengths for $\theta_p$= 28.74$^{\circ}$. (a) $L=1$ mm, (b) $L=2$ mm, (c) $L=3$ mm.} 
\label{L_variation_1}
\end{figure}

\begin{figure}[t!]
\centering
\includegraphics[scale=1.0]{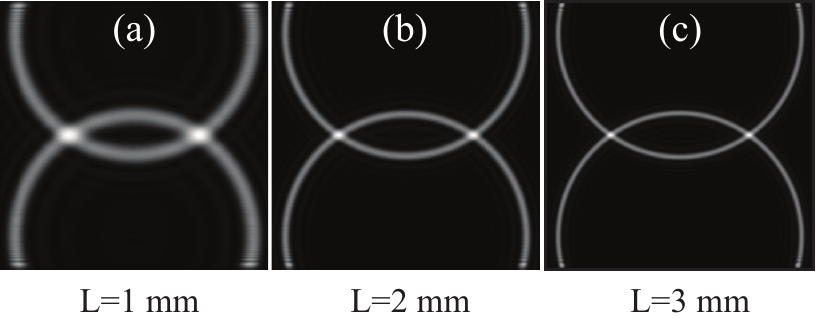}
\caption{Numerically simulated down-conversion intensity patterns for type-II phase matching at different crystals lengths for $\theta_p$= 41.78$^{\circ}$.  (a) $L=1$ mm, (b) $L=2$ mm, (c) $L=3$ mm.}
\label{type_II_41_78_Lvariation_theory}
\end{figure}

The length of the crystal $L$ directly affects the phase matching in  SPDC. Figure \ref{L_variation_1} shows the down-conversion intensity pattern for three different values of the crystal thickness $L$ and for type-I phase matching condition at $\theta_p=28.74^{\circ}$. We find that $L$ does not affect the overall non-collinearity of emission. However, it does affect the width of the annulus, which becomes broader as $L$ becomes thinner. This means that as the crystal becomes thinner the phase-matching becomes more relaxed. We can also see this through equation (\ref{eq:19}) that as $L$ increases the $\mbox{sinc}$ term become narrower. Figure \ref{type_II_41_78_Lvariation_theory} shows the down-conversion intensity pattern for three different values of the crystal thickness $L$ for type-II phase matching condition at $\theta_p=41.78^{\circ}$. Just as in the case of type-I phase matching, we find that the overall effect of increasing $L$ is just the tightening of the phase-matching condition which causes the two annuli to become narrower.

\subsection{Effect of $\theta_p$ on the position-correlation width}

\begin{figure}[h!]
\centering
\includegraphics[scale=0.63]{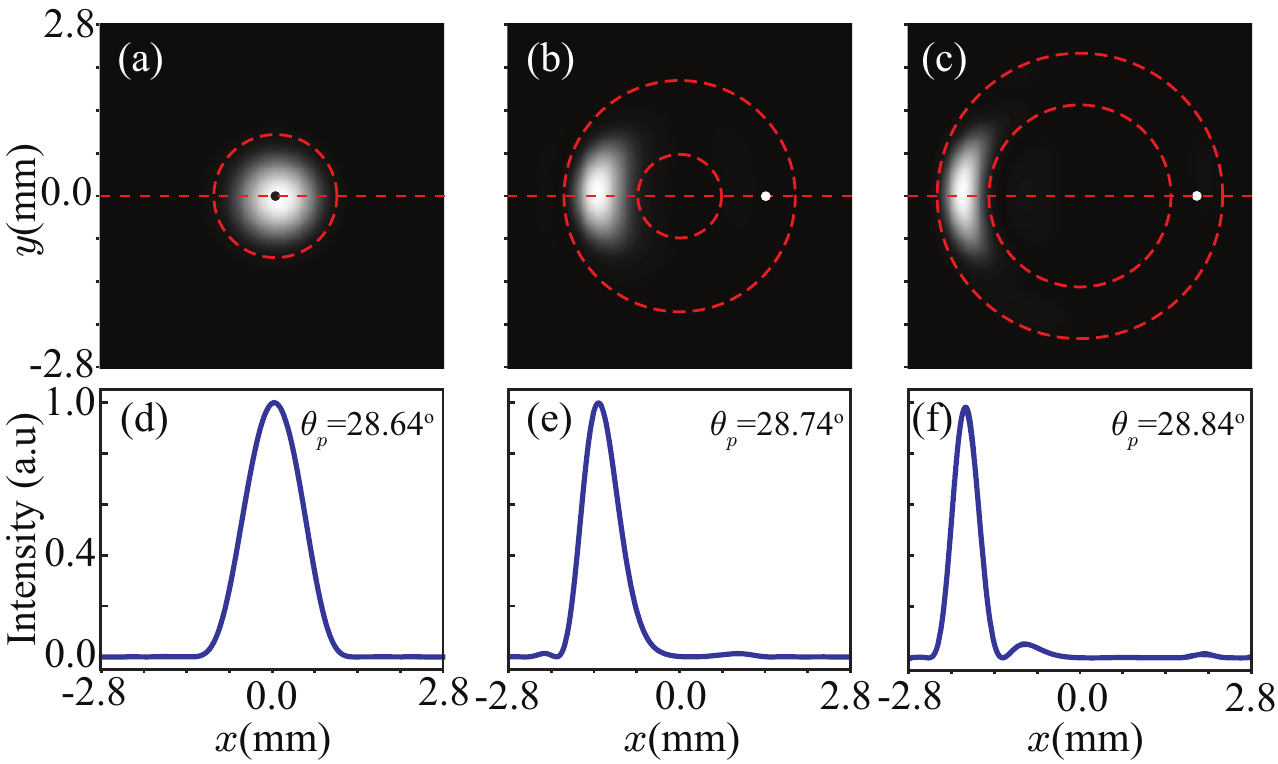}
\caption{Numerical plots of conditional probability of signal photon with fixed idler position 
for (a) $\theta_p=28.64 ^{\circ}$, (b) $\theta_p=28.74 ^{\circ}$, (c) $\theta_p=28.84 ^{\circ}$ the corresponding one dimensional plots  along $x$ direction are shown in (d), (e) and (f) respectively. }
\label{corr_img_all}
\end{figure}
In this subsection, we evaluate the position-correlation width for type-I SPDC for various $\theta_p$ values. Using the expression  given in equation (\ref{conditional_probability_eqn}), we numerically  evaluate the conditional probabilities and plot them in figure \ref{corr_img_all}. In order to get the correct scaling for the pixel size in numerical simulations, we use the fact that the position-correlation width of the signal (idler) photon for a 
fixed position of the idler(signal) photon at any plane is twice the pump beam size at that plane (see Ref. \cite{jha2010pra}). In figure \ref{corr_img_all}, we find that for the collinear case the correlation area forms a circle but as the non-collinearity increases the correlation area no longer remains symmetric. Figures \ref{corr_img_all}(a), \ref{corr_img_all}(b), and \ref{corr_img_all}(c) show the plots of two-dimensional conditional probablility at various $\theta_p$ values. Figures \ref{corr_img_all}(d), \ref{corr_img_all}(e), and \ref{corr_img_all}(f) are the one-dimensional cuts along the $x$-direction. Using these one dimensional plots, we calculate the correlation width along the $x$ direction and find it to be  $847 \ {\rm \mu m}$, $581 \ {\rm \mu m }$, and $401 \ {\rm \mu m }$, respectively, at the three $\theta_p$ values.

\section{Two-photon wavefunction in the orbital angular momentum basis }\label{angular_spectrum}

We have so far mostly discussed the two-photon wavefunctions in the position and momentum bases. The other set of bases that have more recently become very important for quantum information is the angular-position and orbital-angular-momentum (OAM) basis. There are several quantum information applications that have been proposed based on the OAM of photons. 

\subsection{Introduction to the orbital angular momentum of a photon}

It is known that Laguerre-Gaussian modes are the solutions to the paraxial Helmholtz equation. The Laguerre-Gaussian modes form a family of orthogonal modes that have a well defined orbital angular momentum. The amplitude of a Laguerre-Gaussian (LG) mode has an azimuthal phase dependence $e^{-il\phi}$, where $l$ is called the azimuthal mode index. Allen {\it et al.} showed that a Laguerre-Gaussian mode with index $l$ possesses an orbital angular momentum of $l\hbar$
per photon  \cite{allen1992pra}. Thus, within quantum mechanics, $l$ has the
interpretation that a field with a Laguerre-Gaussian amplitude distribution carries an orbital angular momentum of $l\hbar$ per
$\hbar\omega$ photon energy \cite{allen1992pra}. The index $p$ represents the radial mode index and decides how the field amplitude gets distributed in the radial direction.

\subsection{Derivation of the two-photon wavefunction}

The two photon wavefunction generated by SPDC process can be written in the in LG basis as 
\begin{equation}
\qquad |\psi_{\rm tp}\rangle = \sum_{l_s,p_s}\sum_{l_i,p_i}C^{l_s, p_s}_{l_i,p_i} |l_s, p_s\rangle_s |l_i, p_i\rangle_i,
\end{equation} 
where $|l_s, p_s\rangle_s$ represents the state of a signal photon in the LG mode having orbital angular mode index $l_s$ and the radial mode index $p_s$. Using the momentum basis representation of LG mode, that is, $\langle {\bm q}| l , p \rangle = LG^{l}_{p}({\bm q}) $ \cite{torres2003pra} in equation (\ref{psitp}), we write $ C^{l_s, p_s}_{l_i,p_i}$ as \cite{miatto2011pra, yao2011njp}
\begin{eqnarray}
 C^{l_s, p_s}_{l_i,p_i} =  A\int\!\!\!\!\int d^2 {\bm q}_s  d^2 {\bm q}_i  V({\bm q}_s+{\bm q}_i) \nonumber \\ 
\qquad   \times\Phi({\bm q}_s,{\bm q}_i) \left[ LG^{l_s}_{p_s}({\bm q}_s)\right]^{*}\left[ LG^{l_i}_{p_i}({\bm q}_i)\right]^{*}.
\end{eqnarray}
Now using the polar coordinate in transverse momentum space, that is ${\bm q}_s =(q_{sx}, q_{sy})=(\rho_s {\rm cos} \phi_s , \rho_s {\rm sin} \phi_s)$, ${\bm q}_i =(q_{ix},q_{iy})=(\rho_i {\rm cos} \phi_i , \rho_i {\rm sin} \phi_i)$, $d {\bm q}_s = \rho_s d \rho_s d \phi_s$ and $d {\bm q}_i = \rho_i d \rho_i d \phi_i$ we can write $C^{l_s, p_s}_{l_i,p_i} $ as 
\begin{eqnarray}
 C^{l_s, p_s}_{l_i,p_i}= \int\!\!\!\!\int_{0}^{\infty}\!\!\!\!\int\!\!\!\!\int_{-\pi}^{\pi}  V(\rho_s, \rho_i,\phi_s,\phi_i)\Phi(\rho_s, \rho_i,\phi_s,\phi_i) \nonumber \\ 
  \times\left[ LG^{l_s}_{p_s}(\rho_s,\phi_s)\right]^{*} \left[ LG^{l_i}_{p_i}(\rho_i,\phi_i)\right]^{*}\rho_s\rho_i d\rho_s d \rho_id \phi_s d \phi_i.\nonumber\\ \label{c_ls_li_polar}
\end{eqnarray}

\subsection{Calculation of the angular Schmidt spectrum}

The probability of detecting signal and idler photons with OAM values $l_s \hbar$ and $l_i \hbar$ respectively, can be represented as 
\begin{equation}
 P^{l_s}_{l_i} = \sum^{\infty}_{p_s=0}\sum^{\infty}_{p_i=0} |C^{l_s,p_s}_{l_i,p_i}|^2.
\end{equation}   
Now using equation (\ref{c_ls_li_polar}), we can get 
\begin{eqnarray}
 P^{l_s}_{l_i} = \sum^{\infty}_{p_s=0}\sum^{\infty}_{p_i=0} \int\!\!\!\!\int\!\!\!\! \int\!\!\!\! \int_{0}^{\infty}\!\!\!\!\int\!\!\!\!\int\!\!\!\! \int\!\!\!\! \int_{-\pi}^{\pi} V(\rho_s, \rho_i,\phi_s,\phi_i) \nonumber \\ 
  \times V^{*}(\rho'_s, \rho'_i,\phi'_s,\phi'_i)\Phi(\rho_s, \rho_i,\phi_s,\phi_i) \Phi^*(\rho'_s, \rho'_i,\phi'_s,\phi'_i)\nonumber \\
 \times  \left[ LG^{l_s}_{p_s}(\rho_s)\right]^{*}\left[ LG^{l_s}_{p_s}(\rho'_s)\right]\left[ LG^{l_i}_{p_i}(\rho_i)\right]^{*}\left[ LG^{l_i}_{p_i}(\rho'_i)\right] \nonumber \\
\qquad \qquad  \times  e^{i l_s (\phi'_s - \phi_s)+ il_i(\phi'_i - \phi_i)} \nonumber \\
\qquad  \times \rho_s \rho'_s \rho_i \rho'_i d\rho_s d\rho'_s d \rho_i d \rho'_i d \phi_s d \phi'_s d \phi_i d \phi'_i,
\end{eqnarray}
where we have used the relations $LG^{l_s}_{p_s}(\rho_s,\phi_s) = LG^{l_s}_{p_s}(\rho_s)$  $ e^{i l_s \phi_s}$  etc. Using the identity shown in Ref. \cite{jha2011pra}, $ \sum^{\infty}_{p=0}(LG)^{l}_{p}(\rho)(LG)^{*l}_{p}(\rho')= \frac{1}{\pi}\delta(\rho^2 - \rho'^2)$,  in the above equation, we get 
\begin{eqnarray}
 P^{l_s}_{l_i} =  \frac{1}{4 \pi^2} \int\!\!\!\!\int_{0}^{\infty}\!\!\!\!\int\!\!\!\!\int\!\!\!\! \int\!\!\!\! \int_{-\pi}^{\pi} V(\rho_s, \rho_i,\phi_s,\phi_i) V^{*}(\rho_s, \rho_i,\phi'_s,\phi'_i)\nonumber \\ 
  \times \Phi(\rho_s, \rho_i,\phi_s,\phi_i)\Phi^*(\rho_s, \rho_i,\phi'_s,\phi'_i) e^{i l_s (\phi'_s - \phi_s)+ il_i(\phi'_i - \phi_i)} \nonumber \\
 \qquad \qquad \qquad \times \rho_s \rho_i d\rho_s  d \rho_i  d \phi_s d \phi'_s d \phi_i d \phi'_i,
\end{eqnarray} 
which essentially can be represented as  
\begin{eqnarray}
 P^{l_s}_{l_i} =  \frac{1}{4 \pi^2} \int\!\!\!\!\int_{0}^{\infty}\rho_s \rho_i  \Bigg|\int\!\!\!\!\int_{-\pi}^{\pi} V(\rho_s, \rho_i,\phi_s,\phi_i) \nonumber \\ 
 \times\Phi(\rho_s, \rho_i,\phi_s,\phi_i)e^{-i l_s\phi_s- il_i\phi_i} d \phi_s  d \phi_i \Bigg|^2 d\rho_s  d \rho_i  ,
\end{eqnarray} 
Considering the conservation of orbital angular momentum in the SPDC process, we can show that $l_s = -l_i = l$ for Gaussian pump beam \cite{mair2001nat}. In this situation, the corresponding probability $P^{l}_{-l}$ is the angular Schmidt spectrum $S_l$, which can be written as
\begin{eqnarray}
S_l = \frac{1}{4 \pi^2} \int\!\!\!\!\int_{0}^{\infty}\rho_s \rho_i \Bigg|\int\!\!\!\!\int_{-\pi}^{\pi} V(\rho_s, \rho_i,\phi_s,\phi_i) \nonumber \\ 
\times\Phi(\rho_s, \rho_i,\phi_s,\phi_i) e^{-i l(\phi_s - \phi_i)} d \phi_s  d \phi_i \Bigg|^2  d\rho_s  d \rho_i .
\end{eqnarray}
The width of this spectrum depends on $\theta_p$ and hence depends on the rotation of the crystal. This width is characterized by the angular Schmidt number $K_{\alpha}= 1/ \sum_{l}S^{2}_l$ \cite{kulkarni2018pra}. 

\section{Conclusion}\label{conclusion}

Spontaneous parametric down-conversion (SPDC) is the most widely used process for generating photon-pairs entangled in various degrees of freedom such as position-momentum, energy-time, polarization, and orbital angular momentum. In SPDC, a pump photon interacts with a nonlinear optical crystal and splits into two entangled photons called the signal and idler photons. The SPDC process has been studied extensively in the past few decades for various pump and crystal configurations, and the photon pairs produced by SPDC have been used in numerous experimental studies on quantum entanglement and entanglement-based real-world quantum-information applications. In this tutorial article, we have presented a thorough study of phase matching in $\beta$-barium borate (BBO) crystals and have thereby studied the entangled photon generation by SPDC. We have discussed in details the effects on phase-matching due to various crystal and pump parameters such as the length of the crystal, the angle between the optic axis of the crystal and the pump propagation direction, the pump incidence angle on the crystal surface, the refraction at the crystal surfaces, and the pump propagation direction inside the crystal. We have then presented theoretical and experimental results illustrating how various phase matching conditions affect the generation of photon pairs by SPDC. We have also highlighted a few results that could be very useful for experimental investigations with SPDC photons. Finally, we have derived the two-photon wavefunction in the orbital angular momentum basis and have calculated the two-photon angular Schmidt spectrum.

\section*{Acknowledgments}

We acknowledge very fruitful discussions with P. H. Souto Ribeiro and C. H. Monken. We further acknowledge financial support through the research grant no. EMR/2015/001931 from the Science and Engineering Research Board (SERB), Department of Science and Technology, Government of India and through the research grant no. DST/ICPS/QuST/Theme-1/2019 from the Department of Science and Technology, Government of India. SK thanks the University Grant Commission (UGC), Government of India for financial support. NM acknowledges Indian Institute of Technology Kanpur for the postdoctoral fellowship.

\section*{References}

\bibliography{spdc_refnew}
\bibliographystyle{unsrt}
\end{document}